%% file: main.tex
\documentclass{aa}

\usepackage{graphicx}
\usepackage{booktabs}
\usepackage{threeparttable}
\usepackage{threeparttablex}
\usepackage{longtable}
\usepackage{xtab}
\usepackage{xcolor}
\usepackage{array}
\defcitealias{wang2014b}{W14}
\defcitealias{yang2007}{Y07}
\usepackage{txfonts}
\begin{document}

   \title{LoTSS jellyfish galaxies: I. Radio tails in low redshift clusters}


   \author{I.D. Roberts
          \inst{1}
          \and
          R.J. van Weeren\inst{1}
          \and
          S.L. McGee\inst{2}
          \and
          A. Botteon\inst{1}
          \and
          A. Drabent\inst{3}
          \and
          A. Ignesti\inst{4}
          \and
          H.J.A. Rottgering\inst{1}
          \and
          T.W. Shimwell\inst{1,5}
          \and
          C. Tasse\inst{6,7}
          }

   \institute{Leiden Observatory, Leiden University, PO Box 9513, 2300 RA
Leiden, The Netherlands\\
              \email{iroberts@strw.leidenuniv.nl}
        \and
            University of Birmingham School of Physics and Astronomy, Edgbaston, Birmingham, UK
        \and
            Th\"uringer Landessternwarte, Sternwarte 5, D-07778 Tautenburg, Germany
        \and
            INAF- Osservatorio astronomico di Padova, Vicolo Osservatorio 5, IT-35122 Padova, Italy
        \and
            ASTRON, the Netherlands Institute for Radio Astronomy, Postbus 2, 7990 AA, Dwingeloo, The Netherlands
        \and
            GEPI \& USN, Observatoire de Paris, CNRS, Universit\'e Paris Diderot, 5 place Jules Janssen, 92190 Meudon, France
        \and
            Centre for Radio Astronomy Techniques and Technologies, Department of Physics and Electronics, Rhodes University, Grahamstown 6140, South Africa}

   \date{Received September 15, 1996; accepted March 16, 1997}


  \abstract
   {The cluster environment has a strong impact on galaxy star formation, as seen by the fact that clusters host proportionally more red, passive galaxies relative to the field.  Ram pressure stripping may drive this environmental quenching by directly stripping cold gas from galactic disks.  In some cases, ram pressure stripping gives rise to `jellyfish galaxies', observed with clear `tentacles' of stripped gas extending beyond the optical extent of the galaxy.}
   {In this paper we present a large sample of jellyfish galaxies in low redshift clusters ($z<0.05$), identified through 120-168 MHz radio continuum from the LOFAR Two-metre Sky Survey (LoTSS).}
   {From a parent sample of 29 X-ray-detected SDSS galaxy clusters and their spectroscopic members, we visually identify 95 star-forming, LoTSS jellyfish galaxies with 144 MHz radio tails. Star formation rates (SFRs) and stellar masses are obtained for all galaxies from SED fits. For each jellyfish galaxy we determine the tail orientation with respect to the cluster centre and quantify the prominence of the radio tails with the 144 MHz shape asymmetry.}
   {After carefully accounting for redshift-dependent selection effects, we find that the frequency of jellyfish galaxies is relatively constant from cluster to cluster. LoTSS jellyfish galaxies are preferentially found at small clustercentric radius and large velocity offsets within their host clusters and have radio tails that are oriented away from the cluster centre.  These galaxies also show enhanced star formation, relative to both `normal' cluster galaxies and isolated field galaxies, but generally fall within the scatter of the $L_\mathrm{144MHz} - \mathrm{SFR}$ relation.}
   {The properties of the LoTSS jellyfish galaxies identified in this work are fully consistent with expectations from ram pressure stripping.  This large sample of jellyfish galaxies will be valuable for further constraining ram pressure stripping and star formation quenching in nearby galaxy clusters.  We show that LOFAR is a powerful instrument for identifying ram pressure stripped galaxies across extremely wide fields.  Moving forward we will push the search for jellyfish galaxies beyond this initial cluster sample, including a comprehensive survey of the galaxy group regime.}


   \maketitle
%

\section{Introduction} \label{sec:intro}

Satellite galaxies in clusters are significantly more likely to be red and passive relative to similar mass galaxies in the field.  This fact is known as the environmental quenching of galaxy star formation, which has been firmly established in the nearby Universe \citep[e.g.][]{dressler1980,blanton2009,peng2010,wetzel2012,haines2015,roberts2019} and also observed out to $z \sim 1$ and beyond \citep[e.g.][]{muzzin2014,foltz2018,ji2018,old2020}.  Increasingly, ram pressure stripping (RPS) is invoked as a physical driver of this environmental quenching.  As galaxies orbit through the cluster potential, the interaction with the dense intracluster medium (ICM), the strength of which scales as $\rho_\mathrm{ICM} \times v^2$ \citep[e.g.][]{gunn1972} where $\rho_\mathrm{ICM}$ is the density of the ICM and $v$ is the galaxy velocity relative to the ICM, can directly strip gas out of the disk leaving behind a wake of material trailing the galaxy.  If enough cold, star-forming gas is removed via this process, RPS will lead to rapid galaxy quenching in dense environments.  Even if it is just the diffuse atomic gas that is efficiently stripped, this can still lead to star formation quenching, albeit over longer timescales, as the galaxy is exhausting its remaining molecular gas reserves.  Despite RPS being a mechanism to quench galaxies \citep[e.g.][]{gunn1972,quilis2000,roediger2005,bekki2009,steinhauser2016,roberts2019}, star formation may be briefly enhanced during the early stages of the ram pressure interaction, as predicted by simulations \citep[e.g.][]{bekki2003,steinhauser2012,troncoso-iribarren2016,ramos-martinez2018,troncoso-iribarren2020} and seen in observations \citep[e.g.][]{gavazzi1985,poggianti2016,vulcani2018b,roberts2020}.  This star formation enhancement should occur prior to substantial stripping of disk gas and is likely a result of compression in the interstellar medium (ISM). Ram pressure stripping strongly impacts galaxy star formation in clusters; therefore constraining the properties of galaxies experiencing RPS is a key ingredient to understanding the evolution of galaxies in dense environments.
\par
The high ICM densities and large velocity dispersions in massive galaxy clusters are conducive to strong RPS, and therefore clusters are excellent laboratories to study this process in detail.  To do so, one must first identify which cluster galaxies are currently undergoing RPS or which cluster galaxies have experienced strong RPS in the past.  While signatures of RPS have been observed in molecular gas \citep[e.g.][]{sivanandam2010,sivanandam2014,vollmer2012,jachym2014,jachym2019,moretti2020}, dust \citep[e.g.][]{crowl2005}, far-ultraviolet \citep[e.g.][]{smith2010,boissier2012,george2018}, and X-rays \citep[e.g.][]{sun2006,sun2010,poggianti2019b}, most commonly \textsc{Hi} or $\mathrm{H\alpha}$ observations are used to identify jellyfish galaxies \citep[e.g.][]{gavazzi2001,yoshida2002,kenney2004,oosterloo2005,chung2007,sun2007,yagi2007,chung2009,yagi2010,kenney2015,poggianti2017,boselli2018}.  Similarly, cluster galaxies that have highly truncated \textsc{Hi} or $\mathrm{H\alpha}$ profiles (relative to the stellar disk) are interpreted as galaxies that have recently been stripped \citep[e.g.][]{vollmer2007,jaffe2018}.  \textsc{Hi} and $\mathrm{H\alpha}$ observations both trace components that are relatively diffuse and can be strongly influenced by RPS.  As a result, these observations reveal clear RPS tails, leading to highly reliable identification of RPS galaxies.  Furthermore, spectroscopic observations provide detailed kinematic information, giving important insight into the stellar and gas dynamics in these highly disturbed galaxies \citep[e.g.][]{fumagalli2014,bellhouse2017,kalita2019}.  Beyond just the $\mathrm{H\alpha}$ line, optical integral field unit spectroscopy probes stellar kinematics as well as ionization mechanisms via optical line ratios.  Spectroscopic observations provide a wealth of information; however, the downside is that this approach requires expensive observations, typically over relatively small fields of view (FOVs).  This makes it difficult to survey entire galaxy clusters, or better yet many galaxy clusters, and instead leads to observations focused on a relatively small number of bright, prominent galaxies.  Narrow-band imaging of $\mathrm{H\alpha}$ \citep[e.g.][]{yagi2010,boselli2018} sacrifices kinematic information but provides much larger FOVs and shorter integration times.  Narrow-band $\mathrm{H\alpha}$ filters are tied to a specific redshift (outside of tunable filters), which does then limit the scope for observations over multiple galaxy clusters.
\par
An alternative approach is to identify galaxies likely undergoing RPS according to their rest-frame optical morphologies \citep[e.g.][]{mcpartland2016,poggianti2016,roberts2020}.  The biggest advantage to this approach is the wide-field imaging capabilities of many telescopes, allowing for entire clusters to be efficiently imaged, sometimes in a single pointing.  Furthermore, broadband imaging is far less observationally expensive than spectroscopy.  Therefore, imaging permits a more efficient search for RPS in clusters, which can be applied over a range of redshifts. The tradeoff is in the reliability of the RPS galaxies identified in this way.  Rest-frame optical imaging primarily traces the stellar component of the galaxy, which is less perturbed by RPS than \textsc{Hi} or $\mathrm{H\alpha}$.  This means that the RPS features observed in optical imaging are more subtle, making it difficult to conclusively say if a galaxy is undergoing RPS based solely on its optical morphology -- especially since there are other cluster-specific processes that also lead to disturbed optical morphologies, for example tidal effects and/or harassment \citep[e.g.][]{moore1996,mayer2006,chung2007}.
\par
The optimal search for galaxies experiencing RPS will include the best of both by: (a) observing a galaxy component that is readily stripped so that ram pressure tails can be conclusively identified, and (b) surveying a wide-field covering entire clusters so that an unbiased search across all cluster members can be performed.  The LOFAR Two-metre Sky Survey (LoTSS, \citealt{shimwell2017,shimwell2019}) is an excellent survey to accomplish this.  The LoTSS wide area survey will eventually image the entire northern sky at $120-168\,\mathrm{MHz}$, $6''$ resolution, and $100\,\mathrm{\mu Jy\,beam^{-1}}$ noise.  The wide-field nature of LoTSS allows for a uniform survey across many galaxy clusters.  In star-forming galaxies, LoTSS is sensitive to synchrotron emission from cosmic rays accelerated by supernovae.  For star-forming galaxies experiencing RPS, tails of synchrotron emission may be observed as these cosmic rays are stripped from the galaxy disk.  In fact LOFAR has already detected extended emission at 144 MHz around a known jellyfish galaxy in Abell 2626 \citep{poggianti2019,ignesti2020}.  Additionally, previous works have observed RPS features in radio continuum within nearby clusters (primarily around $1.4\,\mathrm{GHz}$, \citealt{gavazzi1987,miller2009,murphy2009,vollmer2009,chen2020,muller2021}).  There is some evidence that RPS tails have steep spectral indices \citep{chen2020,muller2021}, which only furthers the strengths of LoTSS as the low frequencies probed by LOFAR are ideal for detecting steep spectrum sources.
\par
In this work we use LoTSS 144 MHz radio continuum observations to perform a systematic search for galaxies undergoing RPS (`jellyfish galaxies') across 29 X-ray-detected, low redshift ($z<0.05$) galaxy clusters. For the purpose of this work, our definition of jellyfish galaxies are galaxies with stripped tails extending asymmetrically in one direction beyond the optical galaxy disk. We consider all star-forming member galaxies in these clusters and identify $\sim$100 jellyfish galaxies with extended, one-sided radio tails.  Here we introduce the sample and present the general properties of these LoTSS jellyfish galaxies, both relative to other cluster galaxies as well as isolated galaxies in the field.  The paper is outlined as follows.  In Sect.~\ref{sec:methods} we describe the datasets that we use in this work along with the method we take to identify jellyfish galaxies.  In Sect.~\ref{sec:demographics} we describe the general demographics of the jellyfish galaxy sample, including stellar mass and redshift distributions as well as active galactic nuclei (AGN) fractions.  In Sect.~\ref{sec:cluster_fractions} we carefully control for redshift-dependent selection effects in order to explore how the frequency of jellyfish galaxies differ from cluster to cluster.  In Sect.~\ref{sec:PS} we present the positions of LoTSS jellyfish galaxies in projected phase space relative to normal cluster galaxies.  In Sect.~\ref{sec:tail_orientation} we present the orientations of observed 144 MHz tails with respect to the centre of the host cluster for all jellyfish galaxies.  In Sect.~\ref{sec:star_formation} we present the star-forming properties of LoTSS jellyfish galaxies, both relative to the star-forming main sequence and in the context of the radio luminosity - star formation rate relation.  Finally, in Sect.~\ref{sec:conclusions} we present a discussion of our results and highlight the main conclusions from this work.
\par
Throughout this paper we have assumed a $\mathrm{\Lambda}$ cold dark matter cosmology with $\Omega_M=0.3$, $\Omega_\Lambda=0.7$, and $H_0=70\,\mathrm{km\,s^{-1}\,Mpc^{-1}}$.


\section{Data and methods} \label{sec:methods}

\begin{figure}
    \centering
    \includegraphics[width=0.9\columnwidth]{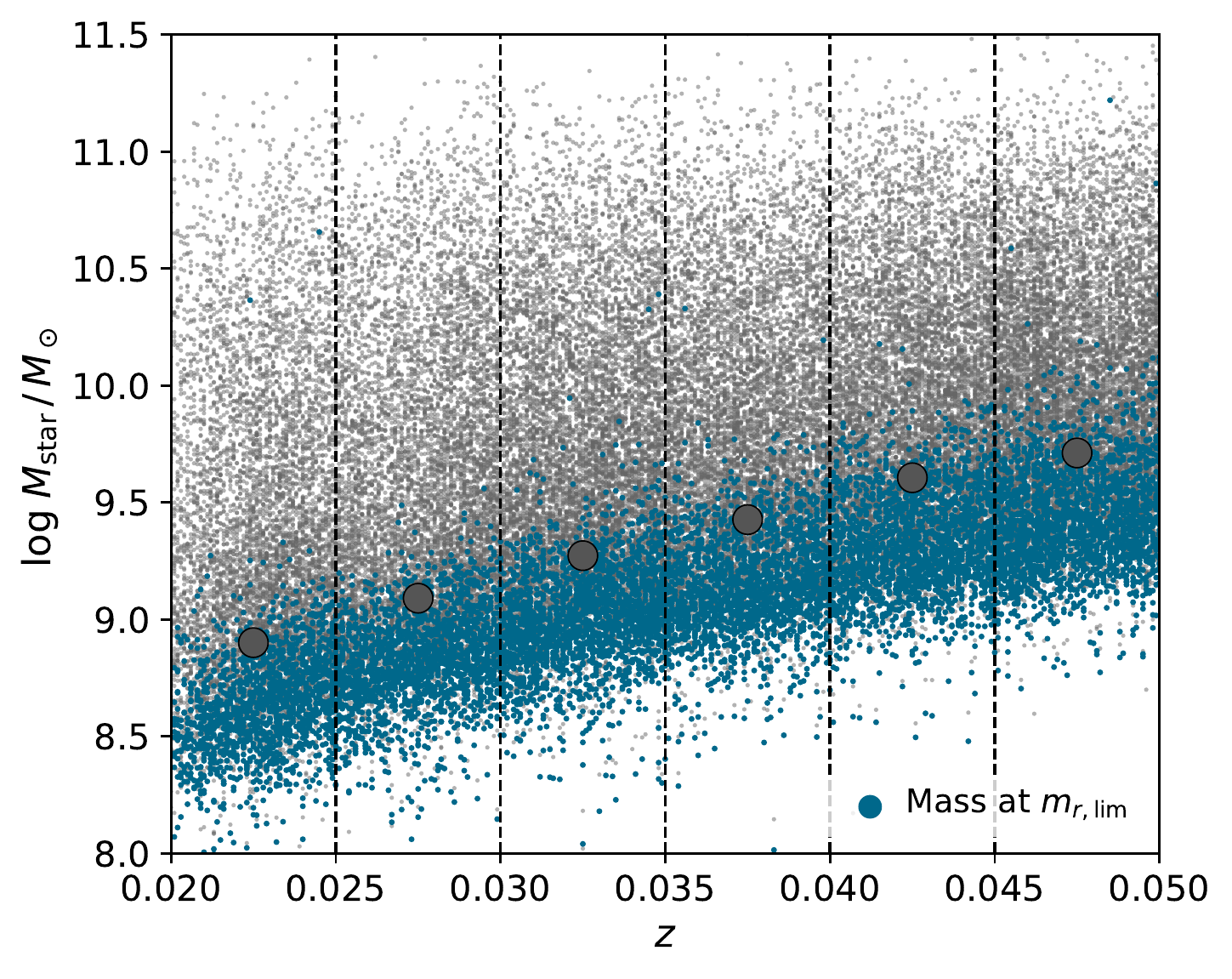}
    \caption{Stellar mass versus redshift for star-forming galaxies in GSWLC-2 \citep{salim2016,salim2018} with $z<0.05$.  Blue points show the limiting stellar mass value for the faintest 20\% of galaxies in each redshift bin and large circles show the stellar mass completeness limits as a function of redshift.}
    \label{fig:mass_lim}
\end{figure}

\begin{figure*}
    \centering
    \includegraphics[width=\textwidth]{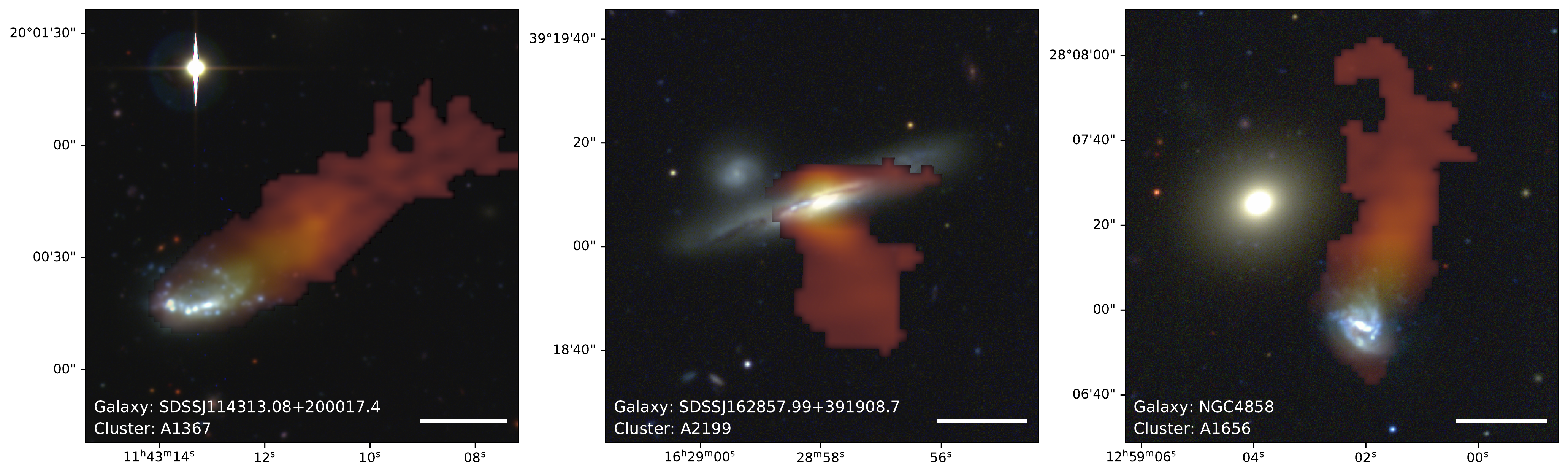}
    \caption{Composite optical (CFHT, RGB) plus radio (LOFAR, 144 MHz) images of three LoTSS jellyfish galaxies in our sample.  The scale bar in each panel corresponds to a physical size of 10 kpc. The radio continuum tail for SDSSJ114313.08+200017.4 was first detected at 1.4 GHz by \citet{gavazzi1985,gavazzi1987}, the radio continuum tail for SDSSJ162857.99+391908.7 is a new detection, and the radio continuum tail for NGC 4858 was first detected at 1.4 GHz by \citet{chen2020}.}
    \label{fig:example_imgs}
\end{figure*}

\subsection{The LOFAR Two-metre Sky Survey (LoTSS)} \label{sec:lotss}

The LOFAR \citep{vanhaarlem2013} Two-metre Sky Survey is an ongoing programme to eventually image the entire northern sky with the LOFAR high-band antenna (HBA) at 120-168 MHz, achieving a typical noise level of $\sim\!100\,\mathrm{\mu Jy / beam}$ and a resolution of $6''$ for most declinations.  We will give a brief description of LoTSS here; however, a full outline of the survey and observing strategy can be found in \citet{shimwell2017}, and the first LoTSS data release (DR1) is described in \citet{shimwell2019}.
\par
LoTSS pointings tile the northern sky with typical separations of $\sim\!2.58^\circ$, with each pointing observed for 8h to reach the desired depth of $\sim\!100\,\mathrm{\mu Jy / beam}$ with direction-dependent calibration \citep{tasse2020}.  With a primary beam FWHM of $\sim\!4^\circ$ at 144 MHz, LoTSS observes the entirety of virtually all low-$z$ clusters in a single pointing.  Our sample includes 29 clusters at $z<0.05$ (see subsequent section) that are observed in LoTSS.  13/29 of these clusters fall within the LoTSS data release 2 (DR2) footprint (\citealt{tasse2020}, Shimwell et al. in prep.), and the other 16/29 have been observed and processed by LoTSS but fall outside of DR2.  For the clusters in our sample, the typical 144 MHz noise levels range from $\sim\!50-250\,\mathrm{\mu Jy / beam}$.  LoTSS noise levels are declination-dependent, with the highest noise levels in our sample corresponding to more equatorial clusters.

\subsection{Cluster sample} \label{sec:cluster_sample}

\begin{table*}
    \centering
    \caption{Cluster sample}
    \begin{threeparttable}
    \begin{tabular}{l c c c c c c c c c}
        \toprule
        Name & \citetalias{wang2014b}\tnote{a} & RA & Dec & $z$ & $M_\mathrm{halo}$\tnote{b} & $N_\mathrm{SDSS}$\tnote{c} & $N_\mathrm{LoTSS}$\tnote{d} & $N_\mathrm{Jellyfish}$\tnote{e} & $N_\mathrm{Jellyfish}^\mathrm{degrade}$\tnote{f} \\
        & Cluster ID & (degrees) & (degrees) & & $(\times 10^{14}\,\mathrm{M_\odot})$ &&&& \\
        \midrule
        A1656 & 1 & 194.8988 & 27.9593 & 0.024 & 7.6 & 265 & 34 & 29 & 8\\
        A2147 & 2 & 240.5709 & 15.9747 & 0.036 & 6.8 & 194 & 18 & 8 & 7 \\
        A2151 & 6 & 241.1492 & 17.7216 & 0.037 & 5.7 & 104 & 23 & 3 & 2 \\
        A2197 & 4 & 246.9214 & 40.9270 & 0.031 & 5.4 & 122 & 29 & 9 & 1 \\
        A2040 & 14 & 228.2123 & 7.4634 & 0.045 & 4.6 & 44 & 3 & 1 & 1 \\
        A2593 & 8 & 351.0837 & 14.5372 & 0.042 & 4.6 & 77 & 14 & 2 & 0 \\
        A1367 & 3 & 176.2709 & 19.6064 & 0.022 & 4.4 & 105 & 7 & 4 & 1 \\
        A2152 & 10 & 241.3716 & 16.4359 & 0.043 & 4.3 & 54 & 10 & 1 & 1 \\
        A2199 & 5 & 247.1631 & 39.5528 & 0.030 & 4.1 & 101 & 19 & 9 & 7 \\
        A1185 & 7 & 167.6785 & 28.6931 & 0.033 & 3.5 & 118 & 24 & 3 & 1 \\
        A168 & 25 & 18.7400 & 0.3113 & 0.045 & 3.1 & 66 & 13 & 3 & 3 \\
        A2052 & 24 & 229.1854 & 7.0864 & 0.035 & 2.8 & 32 & 4 & 1 & 0 \\
        A2107 & 21 & 234.9127 & 21.7827 & 0.041 & 2.7 & 44 & 11 & 3 & 1 \\
        A2063 & 17 & 230.7721 & 8.6092 & 0.035 & 2.6 & 58 & 10 & 2 & 2 \\
        MKW3s & 51 & 230.4037 & 7.7193 & 0.045 & 2.2 & 26 & 2 & 3 & 2 \\
        NGC4065 Grp & 15 & 181.0448 & 20.3509 & 0.024 & 2.0 & 92 & 12 & 0 & 0 \\
        MCXC J1722.2+3042 & 183 & 260.6657 & 30.8805 & 0.046 & 1.9 & 21 & 4 & 0 & 0 \\
        SDSS-C4-DR3 3088 & 73 & 146.8347 & 54.4917 & 0.046 & 1.8 & 17 & 2 & 0 & 0 \\
        MKW8 & 22 & 220.1785 & 3.4654 & 0.027 & 1.7 & 54 & 2 & 0 & 0 \\
        A1314 & 43 & 173.7054 & 49.0776 & 0.033 & 1.6 & 39 & 9 & 2 & 0 \\
        NGC5098 Grp & 71 & 200.0614 & 33.1434 & 0.037 & 1.6 & 35 & 17 & 2 & 0 \\
        W14\_68 & 68 & 245.7615 & 37.9223 & 0.031 & 1.3 & 35 & 9 & 5 & 0 \\
        W14\_88 & 88 & 227.8814 & 4.5175 & 0.037 & 1.3 & 43 & 8 & 1 & 1 \\
        MCXC J1010.2+5429 & 111 & 152.5529 & 54.4864 & 0.046 & 1.3 & 25 & 8 & 0 & 0 \\
        NGC6107 Grp & 45 & 244.3727 & 34.9578 & 0.031 & 1.2 & 46 & 7 & 1 & 0 \\
        A779 & 35 & 139.9453 & 33.7497 & 0.023 & 1.1 & 45 & 8 & 0 & 0 \\
        A1228 & 139 & 170.5129 & 34.3148 & 0.035 & 1.1 & 36 & 9 & 0 & 0 \\
        NGC6338 Grp & 72 & 258.8457 & 57.4112 & 0.029 & 1.1 & 44 & 12 & 3 & 1 \\
        ZwCl 2212+1326 & 90 & 333.7197 & 13.8406 & 0.026 & 1.1 & 26 & 2 & 0 & 0 \\
        \bottomrule
    \end{tabular}
    \begin{tablenotes}
    \item[a] \citet{wang2014b}
    \item[b] Abundance matching from \citet{yang2007}
    \item[c] Number of star-forming SDSS galaxies per cluster
    \item[d] Number of star-forming LoTSS detected galaxies per cluster
    \item[e] Number of LoTSS jellyfish galaxies per cluster
    \item[f] Number of LoTSS jellyfish galaxies per cluster after image degredation (see text)
    \end{tablenotes}
    \end{threeparttable}
    \label{tab:clusters}
\end{table*}

The galaxy clusters in this work were selected from the Sloan Digital Sky Survey (SDSS, \citealt{york2000,abazajian2009}), which has substantial overlap with the LoTSS footprint.  We selected galaxy clusters from the SDSS-ROSAT All Sky Survey (RASS, \citealt{voges1999}) cluster catalogue from \citet{wang2014b}, hereafter \citetalias{wang2014b}.  \citetalias{wang2014b} measure RASS X-ray luminosities around optical SDSS clusters from the \citet{yang2007}, hereafter \citetalias{yang2007}, catalogue.  From the \citetalias{wang2014b} catalogue, we selected all low redshift ($z<0.05$) clusters with X-ray detections that had also been observed by LOFAR at $144\,\mathrm{MHz}$ at the time of writing.  Specifically, we select clusters from the \citetalias{wang2014b} catalogue with redshifts $<0.05$, RASS X-ray detections with S/N $>3$, halo masses $\ge 10^{14}\,h^{-1}\,\mathrm{M_\odot}$, and require that the cluster has been observed by LOFAR at $144\,\mathrm{MHz}$. For the clusters in \citetalias{wang2014b}, we take the corresponding halo masses, $M_\mathrm{halo}$, from the \citetalias{yang2007} catalogue.  These halo masses are computed via abundance matching and we use the masses that are matched according to the characteristic stellar mass for each cluster.  A full description of the abundance matching method is given in Sect. 3.5 in \citetalias{yang2007}, and comparison to mocks shows that this method is able to reproduce halo masses without bias and with a scatter of $\sim\!0.2\,\mathrm{dex}$.
\par
The criteria above yield a final sample of 29 X-ray-detected SDSS clusters with LoTSS observations.  These clusters and their basic properties are listed in Table~\ref{tab:clusters}.  For each cluster we determined the 1D velocity dispersion assuming the following scaling from \citetalias{yang2007}
\begin{equation}
    \sigma = 397.9\,\mathrm{km\,s^{-1}}\,\left(\frac{M_\mathrm{halo}}{10^{14}\,h^{-1}\,\mathrm{M_\odot}}\right)^{0.3214}.
\end{equation}
\noindent
Following \citetalias{yang2007} we also calculated the virial radius for each cluster as
\begin{equation}
    R_{180} = 1.26\,h^{-1}\,\mathrm{Mpc}\,\left(\frac{M_\mathrm{halo}}{10^{14}\,h^{-1}\,\mathrm{M_\odot}}\right)^{1/3}(1+z_\mathrm{cluster})^{-1},
\end{equation}
\noindent where $R_{180}$ is the radius that encloses an average density equal to 180 times the critical mass density of the Universe.

\subsection{Galaxy samples} \label{sec:galaxy_sample}

\subsubsection{Cluster galaxies} \label{sec:cluster_galaxy_sample}

For this sample of clusters we assign member galaxies, and corresponding star formation rates (SFRs) and stellar masses ($M_\mathrm{star}$), with SDSS spectroscopy.  For each cluster we select galaxies with projected separations from the cluster X-ray centre $<\!R_{180}$ and 1D velocity offsets from the cluster redshift $<\!3\sigma$.  This is a relatively loose membership criterion that may include a small number of galaxies that are not formally bound to the cluster, especially galaxies that have both large separations from the cluster centre and large velocity offsets from the cluster redshift.  We opt for this approach to ensure that we do not miss galaxies that have recently started their infall onto the cluster.  For member galaxies we take stellar masses and SFRs from the GALEX-SDSS-WISE catalogue (GSWLC-2, \citealt{salim2016,salim2018}).  These stellar masses and SFRs are determined by fitting galaxy spectral energy distributions (SEDs) with the \textsc{cigale} code \citep{boquien2019}.  The SED fits include GALEX UV photometry, SDSS optical photometry, and the total IR luminosity estimated from templates using WISE mid-IR fluxes.  In this paper we focus on star-forming cluster galaxies, that we define to be galaxies with a specific star formation rate ($\mathrm{sSFR} = \mathrm{SFR} / M_\mathrm{star}$) $>\!10^{-11}\,\mathrm{yr^{-1}}$ \citep{wetzel2013}.  These selections give a total sample of 1968 star-forming galaxies within our sample of clusters.  For the remainder of the paper we will refer to this sample (excluding galaxies identified as jellyfish, see Sect.~\ref{sec:jellyfish}) as `SDSS cluster galaxies'.
\par
For comparison purposes, we also identify a subset of the SDSS cluster galaxies that are detected by LOFAR at 144 MHz, but are not classified as jellyfish galaxies (see Sect.~\ref{sec:jellyfish}).  To do so we cross match the SDSS cluster galaxy sample with the forthcoming LoTSS DR2 source catalogs (see \citealt{williams2019} for a description of previous LoTSS source catalogs), requiring separations between matching galaxies to be no larger than $3''$.  $3''$ is equal to half of the LoTSS FWHM and corresponds to $1.2-2.9\,\mathrm{kpc}$ over the redshift range of our sample.  We note that the majority of matches have separations of $1''$ or less.  This gives a sample of 330 galaxies, that we will refer to as `LoTSS cluster galaxies'.

\subsubsection{Matched field galaxies} \label{sec:field_sample}

For further comparison we also compiled a sample of SDSS isolated field galaxies from \citet{roberts2017}.  This sample was derived from $N=1$ `groups' in the \citetalias{yang2007} group catalogue, in other words, galaxies that were not assigned to a group of multiple galaxies.  The sample of isolated field galaxies consists of all $N=1$ \citetalias{yang2007} galaxies that are separated from their nearest `bright neighbour' by at least $1\,\mathrm{Mpc}$ and $1000\,\mathrm{km\,s^{-1}}$.  Bright neighbours are defined as any galaxy brighter than the SDSS $r$-band absolute magnitude limit at $z=0.05$ (the upper redshift limit of our cluster sample).  We then selected all galaxies from the \citeauthor{roberts2017} field sample with redshifts $<\!0.05$ and with stellar masses and SFRs from the GSWLC-2 catalog.  Finally, we match the stellar mass and redshift distributions of the field sample to the sample of cluster galaxies.  To do so, for each cluster galaxy we randomly selected five field galaxies within $0.1\,\mathrm{dex}$ in stellar mass and $1000\,\mathrm{km\,s^{-1}}$ in redshift to the cluster galaxy.  This gives a final field sample of 10315 unique galaxies that is well matched to the cluster sample in mass and redshift.

\subsubsection{Stellar mass completeness} \label{sec:mass_complete}

The SDSS is a flux limited spectroscopic survey; therefore, the stellar mass completeness of our galaxy sample will be a clear function of redshift.  To evaluate this stellar mass completeness we use all star-forming galaxies ($\mathrm{sSFR} > 10^{-11}\,\mathrm{M_\odot}$) in the GSWLC-2 catalogue \citep{salim2016,salim2018} with $z < 0.05$, and follow \citet{pozzetti2010,weigel2016} by determining the mass, $M_\mathrm{lim}$, that a given galaxy would have if its magnitude was equal and the limiting magnitude of the SDSS spectroscopic survey, $m_{r,\mathrm{lim}}=17.77$ \citep[e.g.][]{abazajian2009}.  At fixed redshift, this limiting mass can be calculated for each galaxy as
\begin{equation}
    \log M_\mathrm{lim} = \log M_\mathrm{star} + 0.4 \times (m_r - m_{r,\mathrm{lim}}),
\end{equation}
\noindent
where $M_\mathrm{star}$ is the galaxy stellar mass and $m_r$ is the observed $r$-band magnitude.  This relation assumes a constant mass-to-light ratio for each galaxy, which is a reasonable assumption given that we are only including star-forming galaxies.
\par
The $M_\mathrm{lim}$ is computed for each galaxy in the sample.  In narrow bands of redshift ($\Delta z = 0.005$) we selected the faintest 20\% of galaxies (in terms of $m_r$) and determined the value of $M_\mathrm{lim}$ below which 90\% of this faint subsample lies.  This value of $M_\mathrm{lim}$ corresponds to our stellar mass completeness limit for each redshift bin \citep{pozzetti2010,weigel2016}.  In Fig.~\ref{fig:mass_lim} we show the stellar mass as a function of redshift for each galaxy (grey points), as well as $M_\mathrm{lim}$ for galaxies in the faintest 20\% (blue points) and our stellar mass completeness limits as a function of redshift (large grey circles).
\par
We opted to apply a stellar mass completeness cut for each individual cluster, corresponding to the cluster redshift.  This means that lower redshift clusters in our sample will have a completeness cut at smaller stellar mass than higher redshift clusters.  This ranges from $M_\mathrm{star}>10^{8.9}\,\mathrm{M_\odot}$ at $z = 0.02-0.025$ to $M_\mathrm{star}>10^{9.7}\,\mathrm{M_\odot}$ at $z = 0.045-0.05$.  As an exception to this, any direct cluster to cluster comparisons were always done using the stellar mass completeness cut corresponding to our highest redshift cluster, $M_\mathrm{star}>10^{9.7}\,\mathrm{M_\odot}$.  This is a conservative choice to ensure that any cluster to cluster comparisons are made across a range in stellar mass that is complete for every cluster in the sample.

\subsection{Identifying jellyfish galaxies} \label{sec:jellyfish}

We identify jellyfish galaxies with by-eye classifications.  For each star-forming cluster galaxy we make $100 \times 100\,\mathrm{kpc}$ cutouts from PanSTARRS $g$-band and LoTSS 144 MHz images.  To classify galaxies we use the $g$-band cutout images with LoTSS 144 MHz contours overlaid.  In these overlays we only show 144 MHz contours that are at, or above, the $2\sigma$ level (similar to \citealt{chen2020}).  The overlay images for all jellyfish galaxies identified in this work are shown in Appendix~\ref{sec:img_appendix}.
\par
We visually identify LoTSS jellyfish galaxies as those star-forming cluster galaxies that show: 144 MHz emission that is resolved and clearly asymmetric with respect to the stellar disk of the galaxy (as traced by the g-band flux).  Specifically, we look for `one-sided' asymmetries such that the classifier is able to confidently assign a tail direction visually.  This excludes objects that have complex, asymmetric radio emission that is not one-sided.  We note that many galaxies in our sample are not detected by LoTSS, and therefore we certainly miss some jellyfish galaxies, particularly at low galaxy masses and/or low star formation rates.  The mass distribution of LoTSS detections compared to our total sample of SDSS cluster galaxies is discussed more thoroughly in Sect.~\ref{sec:mass_redshift}.
\par
Through this procedure we identified 95 jellyfish galaxies with LoTSS radio tails.  These galaxies are distributed through 21/29 clusters, whereas eight clusters in our sample have no identified jellyfish galaxies.  The majority of these radio continuum tails are new detections.  \citet{gavazzi1985,gavazzi1987} originally identified radio continuum tails at 1.4 GHz for two of the galaxies in this sample (in Abell 1367, SDSSJ114313.08+200017.4 and KUG1140+202A). \citet{miller2009} also identified radio continuum tails at 1.4 GHz for two of the Coma Cluster galaxies in our sample (GMP4555 and IC4040).  Finally, \citet{chen2020} presented newly detected radio continuum tails (again at 1.4 GHz) for six of the Coma Cluster galaxies in this sample (NGC4848, NGC4858, D100, GMP2599, Mrk0058, GMP3271).  Otherwise, to the best of our knowledge, the rest of the radio continuum tails presented in this work are new detections.  In Fig.~\ref{fig:example_imgs} we show optical+radio images of three example jellyfish galaxies in our sample.  The optical imaging in Fig.~\ref{fig:example_imgs} is from the MegaCam instrument on the Canada-France-Hawaii Telescope\footnote{https://www.cadc-ccda.hia-iha.nrc-cnrc.gc.ca/en/cfht/} (CFHT).

\subsection{LoTSS asymmetries} \label{sec:144asym}

We measure the `shape asymmetry' \citep{pawlik2016} of the $\mathrm{144\,MHz}$ emission for the identified jellyfish galaxies.  The shape asymmetry is simlar to the commonly used CAS rotational asymmetry \citep{abraham1996,conselice2003}, but instead of being calculated from the flux image it is calculated from a binary detection map for a given source.  In this sense, it is a measure of asymmetry that is not flux-weighted, therefore making it very sensitive to low surface brightness features such as stripped tails.  This asymmetry, $A_S$, is given by
\begin{equation}
    A_S = \frac{\sum | X_0 - X_{180} |}{2\sum |X_0|},
\end{equation}
\noindent
where $X_0$ is the LoTSS segmentation map and $X_{180}$ is the LoTSS segmentation map rotated by $180^\circ$.  We measure $A_S$ with respect to the optical centre of each galaxy.  Segmentation maps are generated using the \texttt{photutils.detect\_sources} function in \textsc{Python} and require sources to have at least 5 connected pixels above a $2\sigma$ threshold.  All segmentation maps are checked manually to ensure that they accurately capture all radio emission from the jellyfish galaxies.  In a small number of cases ($\sim$5\%) the segmentation maps blend two sources together, and for these cases \texttt{photutils.deblend\_sources} deblends the segmentation map at saddle points in the image using watershed segmentation.


\section{Jellyfish galaxy demographics} \label{sec:demographics}

\subsection{Mass and redshift distributions} \label{sec:mass_redshift}

\begin{figure}
    \centering
    \includegraphics[width=0.9\columnwidth]{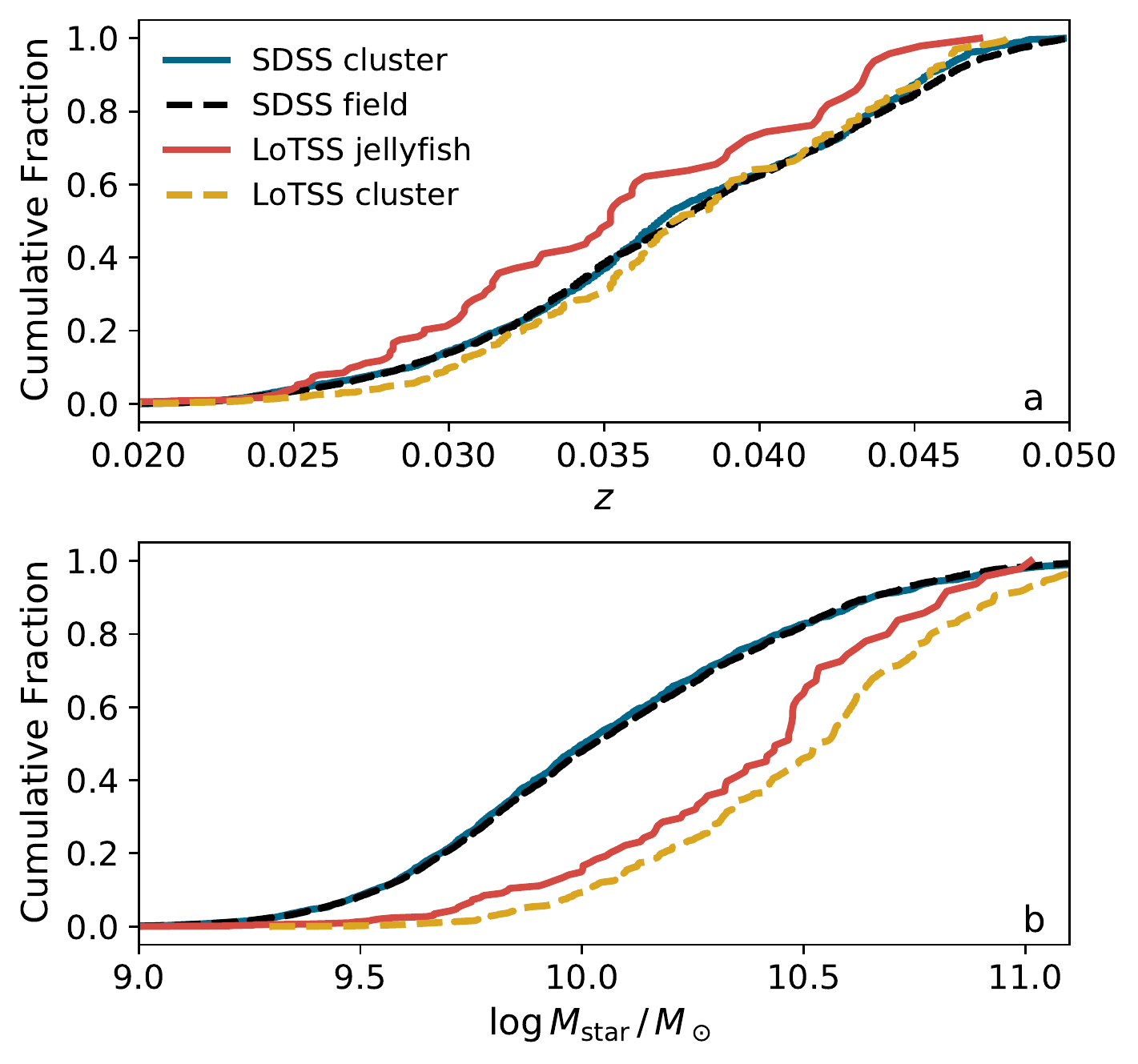}
    \caption{Empirical distribution functions (EDFs) for galaxy redshift (top) and galaxy stellar mass (bottom).  EDFs are shown for SDSS cluster galaxies (blue, solid), SDSS field galaxies (black, dashed), LoTSS jellyfish galaxies (red, solid), and normal star-forming galaxies from the LoTSS DR2 source catalogue (red, dashed).}
    \label{fig:mass_z_dist}
\end{figure}

The sample of cluster or field galaxies with SDSS spectroscopy are subject to different selection functions, in terms of galaxy stellar mass and redshift, than the sample of galaxies detected by LoTSS at 144 MHz.  Since our cluster sample comprises a range of redshifts ($z_\mathrm{cluster} \sim 0.02-0.05$) and galaxy stellar masses, it is important to understand these different selection effects in order to ensure fair comparisons throughout our galaxy samples and between jellyfish galaxy populations from cluster to cluster.
\par
In Fig.~\ref{fig:mass_z_dist}a we plot empirical distribution functions (EDFs) of galaxy redshift for SDSS cluster galaxies (blue, solid), SDSS field galaxies (black, dashed), LoTSS jellyfish galaxies (red, solid), and LoTSS cluster galaxies (gold, dashed).  As discussed in Sect.~\ref{sec:field_sample}, the SDSS field sample is explicitly matched to the SDSS cluster sample in terms of redshift (and stellar mass), so the identical distributions between the two in Fig.~\ref{fig:mass_z_dist}a is by construction.  The redshift distribution of LoTSS cluster galaxies is also essentially identical to the SDSS cluster and field samples; however, this is not by construction.  On one hand this is unsurprising as the LoTSS cluster sample is a subset of the SDSS cluster sample, but it does show that galaxies detected by LOFAR are not preferentially found at the low redshift end of our sample.  On the other hand, LoTSS jellyfish galaxies have a distribution that is clearly shifted to low redshifts.  While the galaxies in the LoTSS jellyfish sample would be detected in LoTSS across the entire redshift range of our sample (similar to the LoTSS cluster galaxies), it does become more difficult to identify a galaxy as a jellyfish at the higher redshift end.  This is the result of a combination of factors: (a) while the galaxy as a whole may be detected in LoTSS across the entire redshift range, low surface brightness tails from RPS can fall below the detection limit in higher redshift clusters, and (b) the size of the beam in physical units increases with redshift, making it more difficult to identify asymmetric tails as they become more blurred at higher redshifts.
\par
In Fig.~\ref{fig:mass_z_dist}b we plot EDFs of stellar mass for the same four samples. Again, by construction the SDSS cluster sample and the matched SDSS field sample have identical distributions.  In terms of stellar mass, both LoTSS cluster galaxies and LoTSS jellyfish galaxies have systematically high masses relative to the pure SDSS samples.  This difference is introduced by the sensitivity limit of LoTSS which does not reach the necessary depths to detect the lowest-mass galaxies in the SDSS samples.  As a result of this selection effect, with this sample we are only able to probe jellyfish galaxies with $\log M_\mathrm{star} / M_\odot \gtrsim 9.5-10$ or so.  This does not mean that there are no jellyfish galaxies below these masses, previous studies actually suggest that the frequency of jellyfish galaxies likely increases towards lower masses (\citealt{yun2019}, Roberts et al. submitted), but deeper observations are needed to probe the regime around $10^9\,\mathrm{M_\odot}$.
\par
The distributions in both panels of Fig.~\ref{fig:mass_z_dist} show that any comparisons between the number of jellyfish galaxies from cluster to cluster need to be done with care to account for different selection effects for clusters at different redshifts.  In Sect.~\ref{sec:cluster_fractions} we treat these selection biases in more detail as we explore whether the frequency of jellyfish galaxies differs from cluster to cluster.

\subsection{AGN contamination} \label{sec:agn_fractions}

\begin{figure}
    \centering
    \includegraphics[width=0.9\columnwidth]{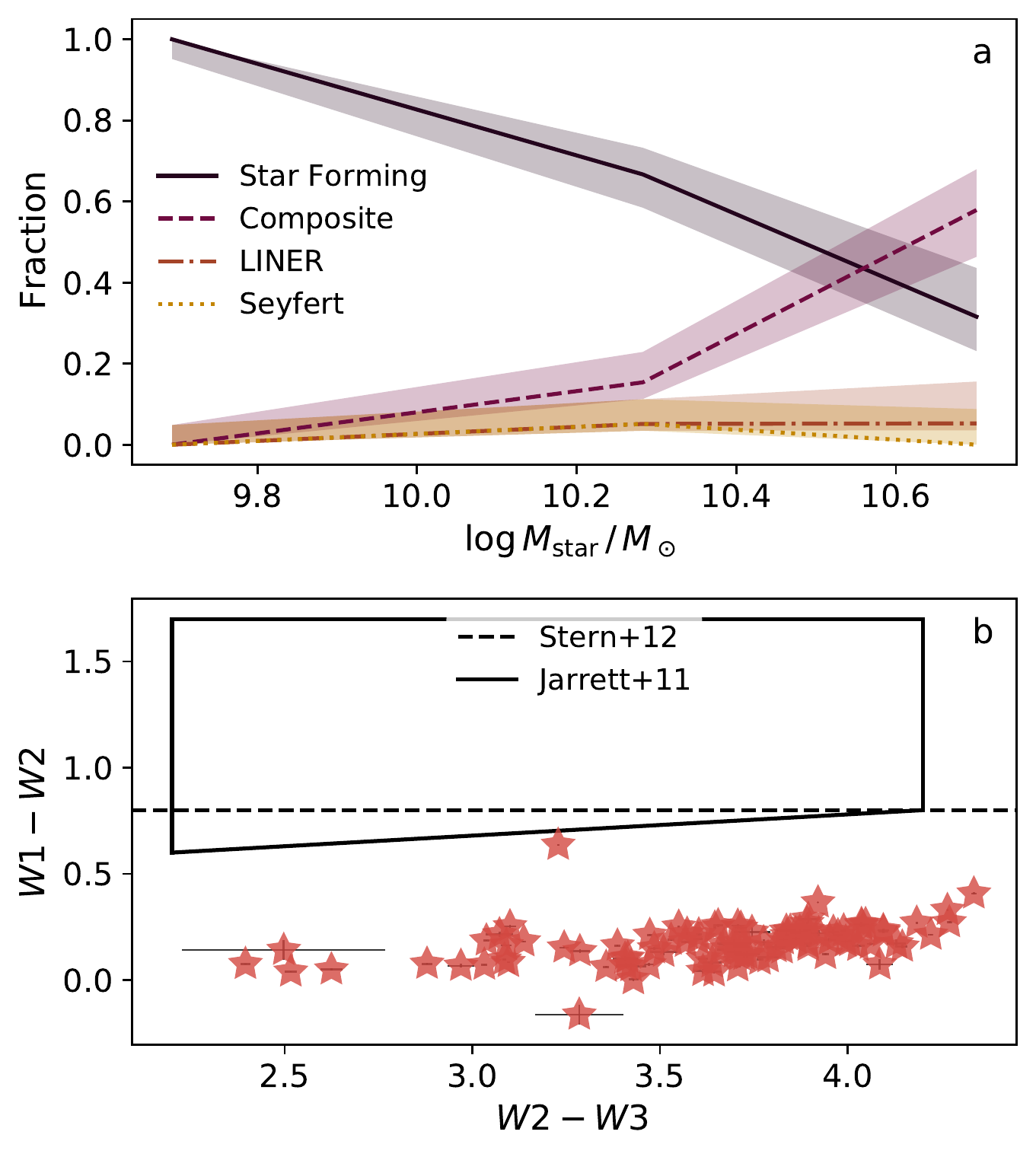}
    \caption{\textit{Top:} Fraction of jellyfish galaxies in the sample according to their BPT classification: star-forming (solid line), composite (dashed line), LINER (dot-dashed line), and Seyfert (dotted line). \textit{Bottom:} WISE colour-colour diagram.  Dashed line shows the dividing line of $W1-W2>0.8$ for AGN classifications from \citet{stern2012} and the solid line shows the AGN region from \citet{jarrett2011}.}
    \label{fig:agn}
\end{figure}

Tailed radio galaxies in clusters typically originate from bent AGN jets \citep[e.g.][]{miley1980,garon2019}.  Given that the search in this work focuses on star-forming galaxies, the source of the 144 MHz tails that we observe is likely not due to AGN.  That said, we can further test for any evidence of AGN emission in these galaxies with optical emission line and mid-IR AGN diagnostics.
\par
In Fig.~\ref{fig:agn}a we plot the fractions of jellyfish galaxies, as a function of galaxy mass, belonging to each of the four classes from the BPT diagram (star-forming, composite, LINER, and Seyfert; \citealt{baldwin1981}).  BPT classifications are from SDSS spectroscopy and are taken from \citet{thomas2013}.  Star-forming, composite, and AGN type emission are differentiated using the \citet{kewley2001} and \citet{kauffmann2003} dividing lines.  The AGN class is further subdivided into Seyfer and LINER using the dividing line from \citet{schawinski2007}.  For all stellar masses, the fraction of galaxies with LINER- or Seyfert-type emission is very small, at most $5$-$10$\%.  Low-mass galaxies are dominated (75-100\%) by star-forming emission line ratios.  Only for the highest-mass galaxies is there a substantial contribution from composite-type emission, suggesting for high-mass galaxies that there may be a blend of star formation and AGN emission contributing to the observed line ratios.  On the whole, 72\% of jellyfish galaxies are classified as star-forming by the BPT diagram and 93\% are classified as either star-forming or composite.  We see no evidence for a prominant population of BPT AGN amongst the jellyfish galaxies as only 2\% of jellyfish are classified as Seyfert-type.  Including both Seyfert- and LINER-type emission only raises this fraction to $7$\%.
\par
Mid-IR photometry is also a useful test for AGN activity \citep[e.g.][]{stern2005}.  In Fig.~\ref{fig:agn}b we plot W1-W2 colour versus W2-W3 colour along with the standard WISE AGN selection regions from \citet{jarrett2011} (solid line) and \citet{stern2012} ($W1-W2>0.8$, dashed line).  The red stars mark the positions of LoTSS jellyfish galaxies in this plane.  WISE Vega magnitudes and errors are taken from the unWISE-SDSS catalogue \citep{lang2016}, and all jellyfish galaxies have $S/N>5$ in all three WISE bands.  The mid-IR colours do not identify any AGN in the sample of jellyfish galaxies, further suggesting that the 144 MHz emission in these galaxies is likely driven by star formation.  Even if we use a less strict AGN selection of $W1-W2>0.5$ \citep[e.g.][]{blecha2018}, that would only identify one AGN galaxy out of the entire jellyfish sample.


\section{Frequency of jellyfish galaxies per cluster} \label{sec:cluster_fractions}

\begin{figure*}
    \centering
    \includegraphics[width=0.9\textwidth]{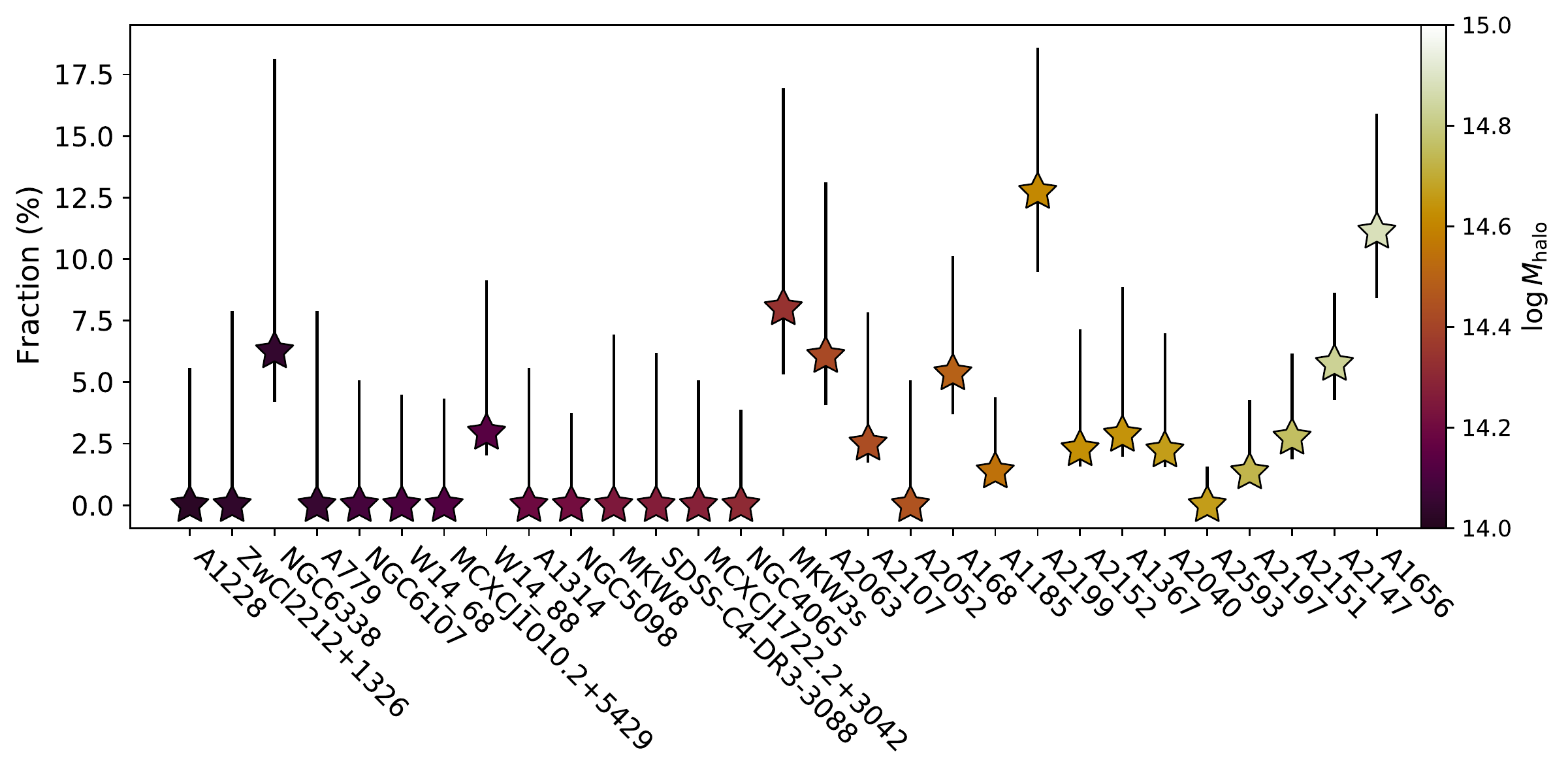}
    \caption{Fraction of star-forming galaxies identified as LoTSS jellyfish for each cluster.  Error bars are $1\sigma$ binomial confidence intervals \citep{cameron2011}.  In order to ensure a fair comparison from cluster to cluster, jellyfish fractions in this plot are measured using only galaxies with $M_\mathrm{star} > 10^{9.7}\,\mathrm{M_\odot}$ and only include jellyfish galaxies identified from the degraded LoTSS imaging.}
    \label{fig:RPfrac_cluster}
\end{figure*}

With the large sample of clusters in this work, we can test whether the frequency of jellyfish galaxies is relatively constant from cluster to cluster, or whether there are certain clusters that are particularly rich in jellyfish.  In order to do this properly, it is crucial to take care of the selection biases in terms of redshift and stellar mass discussed in Sect.~\ref{sec:mass_complete} and Sect.~\ref{sec:mass_redshift}.
\par
We first exclude any galaxies with $M_\mathrm{star} < 10^{9.7}\,\mathrm{M_\odot}$.  This corresponds to the stellar mass completeness limit for our highest redshift clusters ($z=0.045-0.050$) and ensures that we are only considering stellar masses that are complete for every cluster in the sample.  Still, there are further selection effects that need to be accounted for before cluster to cluster comparisons are possible:
\par
First, the variation in sensitivity for 144 MHz images of different clusters.  LOFAR is most sensitive at large declinations ($\delta \gtrsim 20-30^\circ$); therefore, the sensitivity level for equatorial clusters in our sample at $\delta \lesssim 20^\circ$ will be negatively impacted.  Additionally, poor ionosphere conditions or bright sources in the field can introduce high levels of noise, even for observations at high declinations.
\par
Second, we identify jellyfish galaxies according to 144 MHz contours beginning at $2\times$ the rms noise for each $100 \times 100\,\mathrm{kpc}$ cutout image.  Therefore, for lower redshift clusters we are tracing inherently fainter 144 MHz luminosities.  Since tails from RPS are faint, it will be easier to identify these features for galaxies in low redshift clusters, for a fixed rms noise.
\par
Third, the angular size of the LoTSS beam is fixed, that means that the physical scale that the beam traces will change with redshift.  In terms of physical scales, the size of the $6''$ LoTSS beam ranges from $2.7\,\mathrm{kpc}$ to $5.4\,\mathrm{kpc}$ over the redshift range of our cluster sample.  Since galaxies in higher redshift clusters will be more strongly blurred by the LoTSS beam, it becomes more difficult to identify asymmetric features with increasing redshift.
\par
All of these effects need to be addressed before a fair comparison can be made between clusters at different redshifts.  The approach that we take is to degrade the quality of the LoTSS jellyfish images, so that the image quality is relatively constant across our entire cluster sample.  We take the following steps to do so:
\par
(a) For all jellyfish galaxies, 144 MHz images are convolved to a physical resolution of $5.4\,\mathrm{kpc}$.  This corresponds to $6''$ at our highest cluster redshift.
\par
(b) We overlay 144 MHz contours (from the images convolved to $5.4\,\mathrm{kpc}$) on PanSTARRS g-band images for each jellyfish galaxy.  The lowest contour is set to $2 \times 0.25\,\mathrm{mJy/beam} \times (0.046/z)^2$.  $0.25\,\mathrm{mJy/beam}$ corresponds to the maximum rms noise from the sample of jellyfish galaxies, and the factor of $(0.046/z)^2$ ensures that the contour levels are set at $\sim$constant luminosity instead of constant flux.
\par
(c) These `degraded' images are reclassified by-eye in order to check whether the asymmetric radio tails are still visible.
\par
Galaxies where tails are still apparent after this image degradation form a new sample of `degraded' jellyfish galaxies.  39/95 jellyfish galaxies still show radio tails in the degraded images. The remainder of the radio tails were either not detected above the stricter contour levels, were blurred by the larger beam such that they no longer appeared asymmetric, or the jellyfish galaxy fell below the stricter stellar mass completeness cut of $M_\star \ge 10^{9.7}\,\mathrm{M_\odot}$ for all clusters. We note that restricting the sample to $M_\star \ge 10^{9.7}\,\mathrm{M_\odot}$ reduces the number of SDSS cluster galaxies from 1968 to 1056. In Appendix~\ref{sec:img_appendix} we show the original LoTSS images as well as the degraded LoTSS images for all jellyfish galaxies.  This smaller `degraded' jellyfish galaxy sample can now be readily used to compare the frequency of jellyfish galaxies between clusters.  The reduced sample size is a necessity in order to ensure a fair comparison, but unfortunately then limits our ability to make strong statements about the frequency of jellyfish galaxies from cluster to cluster.
\par
In Fig.~\ref{fig:RPfrac_cluster} we plot the fraction of jellyfish galaxies in each cluster.  The fractions for each cluster are measured relative to all star-forming member galaxies, only include galaxies with $M_\mathrm{star} \ge 10^{9.7}\,\mathrm{M_\odot}$ (for stellar mass completeness), and only include jellyfish galaxies from the `degraded' jellyfish galaxy sample.  We stress that these absolute fractions for each cluster are not particularly meaningful since many legitimate jellyfish galaxies are not included due to the image degradation process.  What are meaningful are the difference between fractions for various clusters and this allows us to explore whether certain clusters are more or less jellyfish-rich than average.  Many clusters in Fig.~\ref{fig:RPfrac_cluster} host no jellyfish galaxies from the `degraded' jellyfish sample, including the eight clusters that had no jellyfish from the full sample and also six additional clusters that have jellyfish galaxies identified from the original LoTSS images but either do not have jellyfish galaxies identified from the degraded LoTSS imaging or only have jellyfish galaxies with $M_\mathrm{star} < 10^{9.7}\,\mathrm{M_\odot}$.
\par
On the whole, the clusters in Fig.~\ref{fig:RPfrac_cluster} are essentially consistent with one another, suggesting that the clusters in this sample all host a relatively constant proportion of jellyfish galaxies.  Though again, it is important to note that by homogenizing the selection of jellyfish galaxies through the image degradation, the sample size becomes significantly smaller. We stress that the fact that we find no strong variation from cluster to cluster does not mean that there is no variation, just that we cannot strongly constrain this with the sample at hand.  The only clusters that show any evidence for a departure from the mean value are A2199 and Coma (A1656), but even then those departures are only at the $\sim\!2.5\sigma$ level.


\section{Projected phase space} \label{sec:PS}

\begin{figure*}
    \centering
    \includegraphics[width=0.9\textwidth]{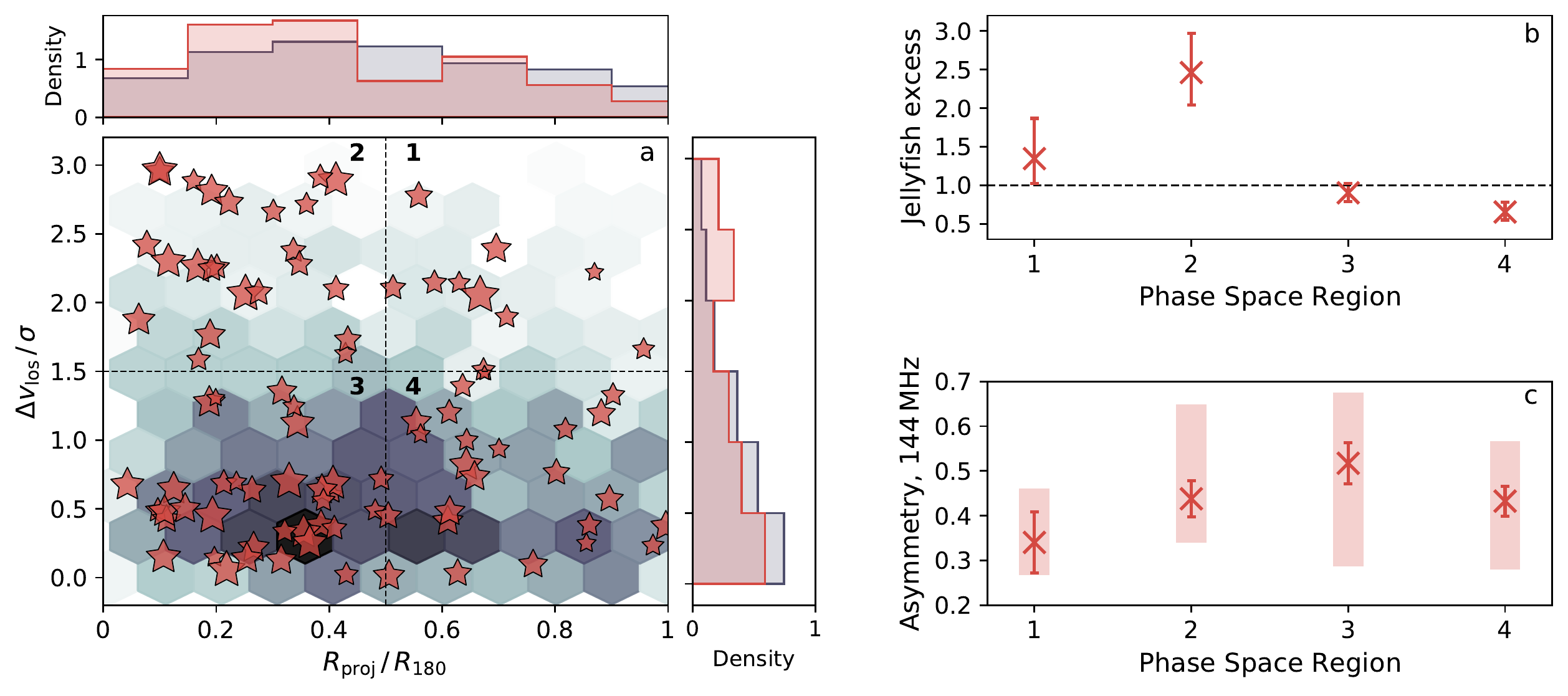}
    \caption{The distribution of SDSS cluster galaxies and LoTSS jellyfish galaxies in projected phase space.  \textit{Left:} Velocity offset versus projected clustercentric radius.  2D histogram shows distribution for star-forming cluster galaxies and red stars correspond to jellyfish galaxies.  We also show the 1D distributions of projected radius and velocity offset over the corresponding axes.  \textit{Right, top:}  Excess of jellyfish galaxies with respect to star-forming cluster galaxies (see text), for the four different regions in projected phase space.  \textit{Right, bottom:} Median 144 MHz asymmetry for each of the four phase space regions.  Error bars show the statistical error on the median and shaded regions show the interquartile range.}
    \label{fig:phase_space}
\end{figure*}

Projected phase space (PPS) diagrams for galaxy clusters are a commonly used tool to constrain cluster accretion and infall history.  Galaxies infall at large velocity offsets (from the cluster systemic velocity) during their first approach to pericentre and after multiple orbits approach the core of phase space at small cluster-centric radius and small velocity offsets (see Fig. 1 in \citealt{rhee2017} for a schematic diagram).  This infall structure in phase space is very clear in simulations where full phase space information is available \citep[e.g.][]{mahajan2011,oman2013,haines2015}.  Projection effects remove some of this clear structure; however, even in projection, galaxies with short times-since infall are generally found at large cluster-centric radius and/or large velocity offsets \citep[e.g.][]{mahajan2011,rhee2017,pasquali2019}.  Given that the strength of ram pressure scales as $\rho_\mathrm{ICM}v^2$, it is expected that jellyfish galaxies should preferentially inhabit regions of PPS at small radius and large velocity offsets.  Indeed, previous observational works have reported high numbers of RPS galaxies in such PPS regions \citep[e.g.][]{yoon2017,jaffe2018,roberts2020}.
\par
In Fig.~\ref{fig:phase_space}a we plot the PPS diagram for the cluster galaxies in our sample.  The background 2D histogram shows the PPS distribution of SDSS cluster galaxies and the red stars correspond to jellyfish galaxies.  The above-axis histograms show the 1D distributions for clustercentric radius and velocity offset, showing that relative to the SDSS cluster galaxies, jellyfish are more commonly found at small clustercentric radius and slightly larger velocity offsets.
\par
We also split PPS into quadrants divided at $R_\mathrm{proj}/R_{180} = 0.5$ and $\Delta v_\mathrm{los}/\sigma = 1.5$, that we label one through four in Fig.~\ref{fig:phase_space}a.  For each quadrant we measure the `excess' of jellyfish galaxies relative to SDSS cluster galaxies.  We define this excess as the ratio between the fraction of jellyfish galaxies in a given quadrant (relative to all jellyfish galaxies) and the fraction of SDSS cluster galaxies in a quadrant (relative to all SDSS cluster galaxies), namely:
\begin{equation}
    \mathrm{Jellyfish\;excess} = \left. \left(\frac{N_\mathrm{jellyfish}^{Q_i}}{N_\mathrm{jellyfish}}\right) \;\right/\; \left(\frac{N_\mathrm{SDSS}^{Q_i}}{N_\mathrm{SDSS}}\right),
\end{equation}
\noindent where $N_\mathrm{jellyfish}^{Q_i}$ is the number of LoTSS jellyfish galaxies in each quadrant, $Q_i$, and $N_\mathrm{jellyfish}$ is the total number of LoTSS jellyfish galaxies, and similarly $N_\mathrm{SDSS}^{Q_i}$ is the number of SDSS cluster galaxies in each quadrant, $Q_i$, and $N_\mathrm{SDSS}$ is the total number of SDSS cluster galaxies.  In Table~\ref{tab:phase_space} we list the number of LoTSS jellyfish galaxies and SDSS cluster galaxies in each of the four phase space quadrants.
\begin{table}
    \centering
    \caption{Galaxy distribution in phase space regions.}
    \begin{tabular}{c c c}
        \toprule
        Phase Space & $N_\mathrm{jellyfish}^{Q_i}$ & $N_\mathrm{SDSS}^{Q_i}$ \\
        Region && \\
        \midrule
         1 & 11 & 154 \\
         2 & 25 & 202 \\
         3 & 41 & 850 \\
         4 & 27 & 762 \\
         \bottomrule
    \end{tabular}
    \label{tab:phase_space}
\end{table}
\par
In Fig.~\ref{fig:phase_space}b we plot this jellyfish galaxy excess as a function of phase space region.  There is a clear excess of jellyfish relative to SDSS cluster galaxies in region 2, that corresponds to small clustercentric radius and large velocity offsets.  This excess is fully consistent with expectations from RPS, as $\rho_\mathrm{ICM}v^2$ will be highest in region 2.  There is also a small deficit of jellyfish galaxies in region 4, corresponding to large clustercentric radius and small velocity offsets.  This region of phase space is difficult to interpret in terms of galaxy infall histories, as it contains a mixture of galaxies just starting their infall onto the cluster as well as `backsplash' galaxies that have already made a pericentric passage and are now approaching their orbital apocentre.  For either case, a small number of jellyfish galaxies may be expected.  For the former, galaxies may have not yet reached the dense part of the ICM required to drive highly asymmetric tails, and for the latter, after a passage through the core of the cluster galaxies may already be mostly stripped.
\par
In Fig.~\ref{fig:phase_space}c we compare the median 144 MHz shape asymmetries in each of the PPS quadrants.  The error bars correspond to statistical errors on the median and the shaded regions cover the interquartile range.  There are no strong trends between 144 MHz asymmetry and PPS region, but the regions corresponding to small clustercentric radii (regions 2 and 3) do have shape asymmetries that scatter to larger values than the regions covering the cluster outskirts.  This is again consistent with prominent, asymmetric tails being primarily driven as galaxies interact with the dense ICM near the centre of the cluster.  While here we have used the non-flux-weighted shape asymmetry parameter, due to its sensitivity to low surface brightness, we note that the same qualitative conclusions arise from using the more standard flux-weighted CAS asymmetry \citep{abraham1996,conselice2003}.


\section{Tail orientations} \label{sec:tail_orientation}

\begin{figure}
    \centering
    \includegraphics[width=0.9\columnwidth]{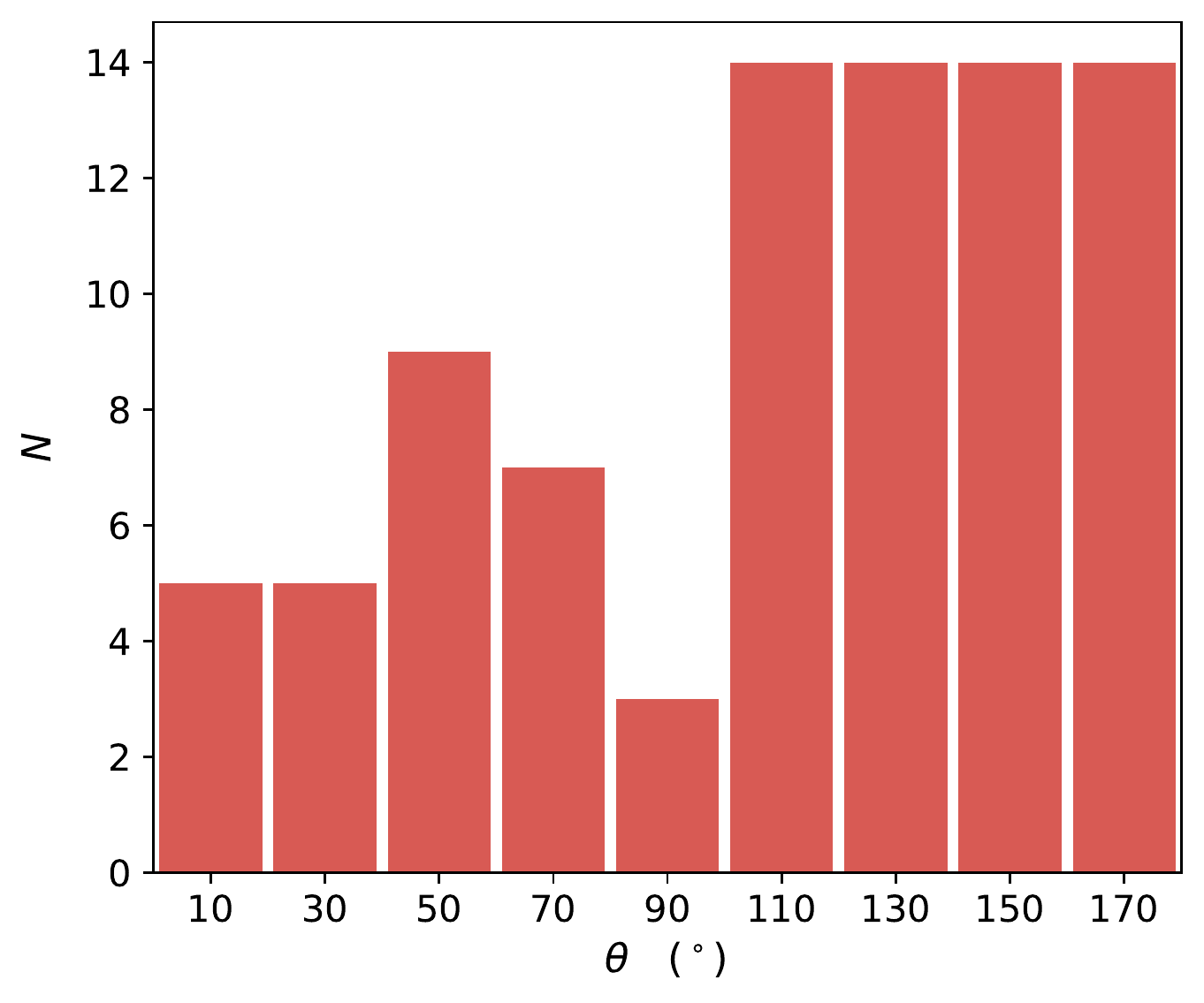}
    \caption{Orientation of LoTSS tails with respect to the X-ray centre of the host cluster.  An angle of $0^\circ$ corresponds to a tail directed towards the cluster centre and an angle of $180^\circ$ corresponds to a tail directed away from the cluster centre.  The method for determining tail orientations is outlined in the main text.}
    \label{fig:tail_orientation}
\end{figure}

Ram pressure stripping makes clear predictions for the orientations of stripped galaxy tails, namely, tails should be observed extending opposite to the direction of motion, which for infalling galaxies on radial orbits will point away from the cluster centre.  This simple interpretation is complicated by variation in galaxy orbital parameters as well as projection effects; however, previous works have reported stripped features extending away from the cluster centre for ram pressure galaxies \citep[e.g.][]{chung2007,smith2010,roberts2020}.
\par
We visually estimate the orientation of the observed 144 MHz tails following the same method as \citet{roberts2020}.  In short, tail directions are estimated from the 144 MHz cutouts as an angle between $0^\circ$ and $360^\circ$, where $0^\circ =$ west and $90^\circ =$ north.  The dot product between the vector along the tail direction and the vector between the galaxy and the cluster X-ray centre gives an orientation angle between the tail direction and the cluster centre. This orientation angle ranges between $0^\circ$ and $180^\circ$, where an angle of $0^\circ$ points directly towards the cluster centre and an angle of $180^\circ$ points directly away from the cluster centre.  These tail direction estimates are shown on the LoTSS jellyfish cutout images in Appendix~\ref{sec:img_appendix}.
\par
In Fig.~\ref{fig:tail_orientation} we show a histogram of the LoTSS tail orientations for the jellyfish galaxies in this work.  If observed tails were randomly distributed then the distribution of orientations would be flat, that clearly is not the case in Fig.~\ref{fig:tail_orientation}.  Instead, Fig.~\ref{fig:tail_orientation} shows a distribution that peaks broadly at angles $>110^\circ$, as expected for tails oriented away from the cluster centre.  These tail orientations are consistent with jellyfish galaxies being primarily stripped on first infall towards the cluster centre, and on relatively radial orbits.  An orientation of $180^\circ$ is only expected for galaxies on purely radial orbits, since cluster galaxy orbits have non-zero tangential components \citep[e.g.][]{biviano2013}, a broader distribution of angles should be expected.


\section{Jellyfish galaxy star formation} \label{sec:star_formation}

\begin{figure*}
    \centering
    \includegraphics[width=0.9\textwidth]{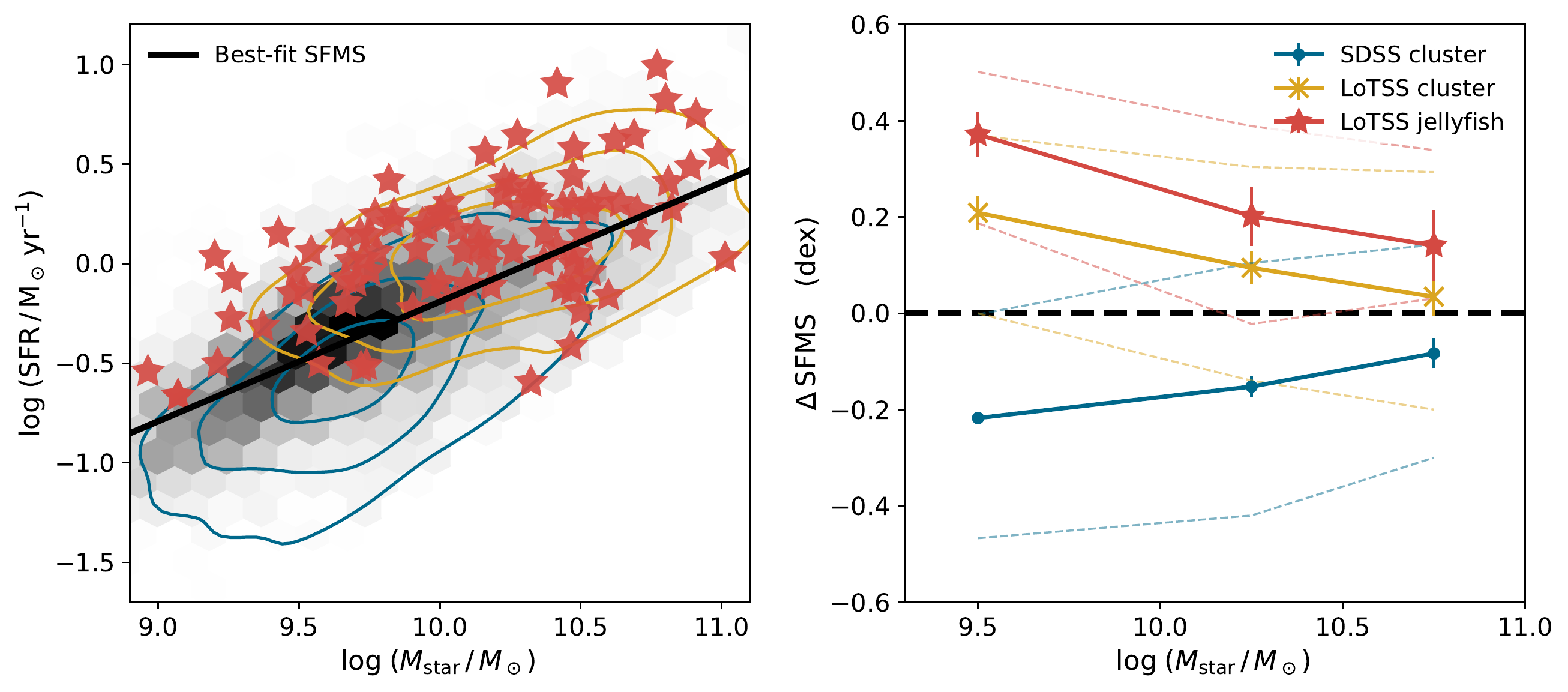}
    \caption{Star-forming properties of SDSS cluster galaxies, LoTSS cluster galaxies, and LoTSS jellyfish galaxies. \textit{Left:}  Star formation rate versus stellar mass for galaxies from the isolated field sample (greyscale), SDSS cluster galaxies (blue contours), LoTSS cluster galaxies (gold contours), and jellyfish galaxies (red stars).  The best-fit star-forming main sequence relation (assuming a power-law relationship) is shown as the solid black line.  \textit{Right:}  Median offset from the SFMS as a function of galaxy mass, for SDSS cluster galaxies (blue circle), LoTSS cluster galaxies (gold cross), and LoTSS jellyfish galaxies (red star).  Error bars correspond to the $1\sigma$ standard error on the median for each stellar mass bin and dashed lines span the interquartile range.}
    \label{fig:SFMS}
\end{figure*}

\begin{figure}
    \centering
    \includegraphics[width=0.9\columnwidth]{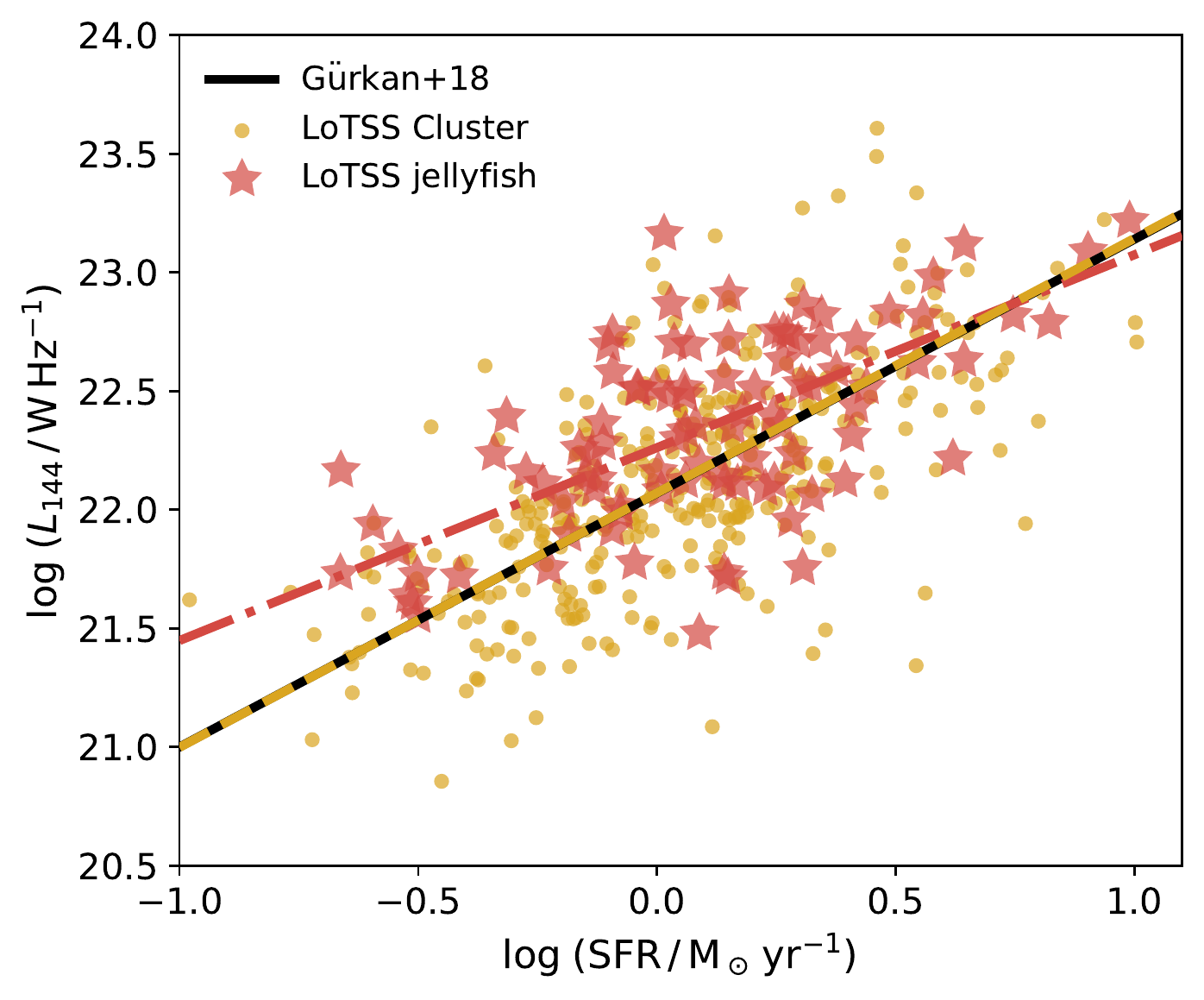}
    \caption{144 MHz luminosity versus star formation rate.  Gold points show the relationship for LoTSS cluster galaxies and red stars correspond to LoTSS jellyfish galaxies.  We also show the best-fit $L_\mathrm{144MHz} - \mathrm{SFR}$ relationship for LoTSS cluster (gold, dash) and LoTSS jellyfish galaxies (red, dot-dash), with shaded regions corresponding to $1\sigma$ bootstrap confidence intervals. The solid black line shows the $L_\mathrm{144MHz} - \mathrm{SFR}$ relationship from \citet{gurkan2018}.}
    \label{fig:radio_cont_sfr}
\end{figure}

The process of RPS should have a strong impact on galaxy star formation.  Not only can RPS quench star formation by removing galaxy gas reserves, but simulations also predict a brief enhancement in star formation, prior to gas removal, due to compression in the ISM through the galaxy-ICM interaction \citep[e.g.][]{steinhauser2012,ramos-martinez2018,troncoso-iribarren2020}.  This picture is supported by recent observations that have shown enhanced levels of star formation in galaxies thought to be undergoing RPS \citep{ebeling2014,poggianti2016,vulcani2018b,roberts2020}.  With the sample in this work we are able to both, test whether LoTSS jellyfish galaxies show enhanced SFRs and explore whether or not jellyfish galaxies fall along the standard 144 MHz luminosity versus SFR relation.

\subsection{The star-forming main sequence} \label{sec:SFMS}

We begin by exploring the position of LoTSS jellyfish galaxies with respect to the star-forming main sequence (SFMS).  In Fig.~\ref{fig:SFMS}a we plot SFR versus stellar mass.  Galaxies from the isolated field sample are shown by the 2D histogram, SDSS cluster galaxies are shown by the blue contours, LoTSS cluster galaxies are shown by the gold contours, and jellyfish galaxies are shown as the red stars.  We overplot the best-fit SFMS relationship for our data, which we obtain by fitting a power-law to the relationship between SFR and stellar mass for galaxies in our isolated field sample.  The fits were performed with the BCES (`bivariate correlated errors and intrinsic scatter', \citealt{akritas1996,nemmen2012}) method with stellar mass as the independent variable, and gives a best-fit SFMS of
\begin{equation}
    \log \mathrm{SFR} = 0.60 \pm 0.01 \times \log M_\star - 6.19 \pm 0.55.
\end{equation}

In Fig.~\ref{fig:SFMS}a we see clear differences between the various galaxy samples.  In terms of stellar mass, both LoTSS cluster galaxies and LoTSS jellyfish galaxies are skewed towards high masses.  This is almost certainly a product of LoTSS not being sensitive to the very low-mass galaxies, as was discussed in Sect.~\ref{sec:mass_redshift}.  In terms of star formation rate, SDSS cluster galaxies skew slightly below the SFMS, LoTSS cluster galaxies mostly straddle the SFMS (though are slightly skewed to high SFRs), and LoTSS jellyfish galaxies lie systematically above the SFMS.  Eighty per cent of jellyfish galaxies are above the SFMS.
\par
The qualitative observations are confirmed quantitatively in Fig.~\ref{fig:SFMS}b, where we plot the offset from the SFMS as a function of stellar mass.  SDSS cluster galaxies (blue) fall below the SFMS by roughly 0.2 dex, with no strong trend with stellar mass.  LoTSS cluster galaxies fall near the SFMS, but do show a small positive offset especially at intermediate and low masses.  The difference between LoTSS cluster galaxies and SDSS cluster galaxies is being driven by selection effects.  The luminosity at 144 MHz is a strong function of star formation activity \citep{gurkan2018,smith2020}; therefore, the galaxies in our sample that are detected by LOFAR will be biased towards high SFRs.  While the same selection effects will also apply to the LoTSS jellyfish galaxies, Fig.~\ref{fig:SFMS}b shows that jellyfish galaxies have enhanced star formation rates even when compared to the LoTSS cluster galaxy sample.  Therefore SFRs for LoTSS jellyfish galaxies are enhanced at a level above and beyond simple selection effects.  This suggests a physical enhancement of star formation in galaxies undergoing RPS in clusters.

\subsection{The radio continuum - SFR relation} \label{sec:radio_cont_sfr}

Previous work has established a tight relationship between 144 MHz luminosity and SFR \citep{gurkan2018,smith2020}.  We now consider whether the LoTSS jellyfish galaxies identified in this work fall along this normal $L_\mathrm{144MHz} - \mathrm{SFR}$ relation.  This potentially gives insight into the source of cosmic rays in the observed tails.  For example, if the cosmic ray emission in jellyfish galaxies is not solely due to star formation then offsets from the $L_\mathrm{144MHz} - \mathrm{SFR}$ relation could be expected.  Conversely, if the cosmic rays in the tails are from star formation within the disk of the galaxy, and then were subsequently stripped, we expect jellyfish galaxies to follow the standard $L_\mathrm{144MHz} - \mathrm{SFR}$ relation as long as the flux from the entire tail is accounted for.
\par
In Fig.~\ref{fig:radio_cont_sfr} we plot the $L_{144} - \mathrm{SFR}$ relationship for the sample from this work.  LoTSS cluster galaxies are shown with the gold points and LoTSS jellyfish galaxies are shown with the red stars.  We also show the best-fit $L_{144} - \mathrm{SFR}$ relation (solid line) published in \citet{gurkan2018}, given by:
\begin{equation*}
\log \mathrm{SFR} = 1.07 \pm 0.01 \times \log L_{144} + 22.07 \pm 0.01.
\end{equation*}
\noindent
Star formation rates in \citet{gurkan2018} are measured via SED fits to optical (SDSS $ugriz$) and IR photometry (\emph{WISE} 3.4, 4.6, 12, and 22$\,\mathrm{\mu m}$ and \emph{Herschel} 100, 160, 250, 350, 500$\,\mathrm{\mu m}$) with the \textsc{MagPhys} code. For LoTSS cluster galaxies, 144 MHz luminosities are taken from the LoTSS DR2 source catalogue, where total flux densities are computed using \textsc{PyBDSF} \citep{mohan2015}.  144 MHz luminosities for jellyfish galaxies are computed by summing emission over the same segmentation maps used to calculate 144 MHz asymmetries.  These segmentation maps have been individually inspected to ensure that they cover all extended emission from the jellyfish galaxies; therefore, calculating luminosities in this way ensures that we are including the diffuse emission from the extended tails.  We note for the jellyfish galaxies the luminosities calculated from the segmentation maps agree well with those from the LoTSS source catalogue, with a median ratio (and $1\sigma$ scatter) between the two luminosities of $1.03_{-0.25}^{+0.34}$.
\par
Fig.~\ref{fig:radio_cont_sfr} shows that the data in this work largely follow the relation found by \citet{gurkan2018}.  There is a slight offset to large radio luminosities for LoTSS jellyfish galaxies, though they generally fall within the scatter of other cluster galaxies.  In Fig.~\ref{fig:radio_cont_sfr} we also show the best-fit power-law relations for the samples in this work, the LoTSS cluster galaxies and the LoTSS jellyfish galaxies.  For LoTSS cluster galaxies we find a relation of:
\begin{equation*}
\log \mathrm{SFR} = 1.07 \pm 0.08 \times \log L_{144} + 22.07 \pm 0.02,
\end{equation*}
\noindent
and for LoTSS jellyfish galaxies we find a relation of:
\begin{equation*}
\log \mathrm{SFR} = 0.82 \pm 0.08 \times \log L_{144} + 22.24 \pm 0.03.
\end{equation*}
\noindent
The fits are performed with the BCES method with stellar mass as the independent variable.  The best-fit relationship for LoTSS cluster galaxies is completely consistent with the \citet{gurkan2018} relation, in fact it returns the same slope and normalization.  For LoTSS jellyfish galaxies, the best-fit relation shows some deviation from the \citet{gurkan2018} relationship, particularly at low SFRs, with the slopes differing at the $\sim\!3\sigma$ level and the normalizations differing at the $\sim\!$5$\sigma$ level. This could suggest that for some jellyfish galaxies there is a secondary source of cosmic rays, outside of star formation within the galaxy disk, contributing to the 144 MHz luminosity.  Potentially, this could be low level star formation within the stripped tail \citep{vulcani2018b}, or also weak AGN emission (for example, from the `Composite' objects in Fig.~\ref{fig:agn}a).  The difference between the best-fit LoTSS jellyfish relation and the \citet{gurkan2018} relation is largest at low SFRs.  For high-SFR galaxies, star formation within the galaxy disk should dominate the 144 MHz luminosity and as such these galaxies should agree with the \citet{gurkan2018} relation.  At the low SFR end there may be more room for these secondary sources to contribute to a more substantial portion of $L_\mathrm{144MHz}$.


\section{Discussion \& conclusions} \label{sec:conclusions}

Here we have presented a sample of $\sim$100 jellyfish galaxies in clusters identified on the basis of radio continuum tails at 144 MHz.  These objects are consistent with being primarily stripped on their first infall towards the cluster centre and show enhanced levels of star formation.  As discussed throughout this paper, both of these facts can be explained by strong ram pressure influencing cluster satellite galaxies shortly after infall.   LoTSS 144 MHz observations have proven a very effective tool for identifying jellyfish galaxies over extremely wide areas.  In the remainder of this discussion we will highlight some of the unique ways in which radio continuum observations can constrain the process of RPS moving forward, as well as the prospects for detecting more jellyfish galaxies with LoTSS over a range of environments.

\subsection{Tracing ram pressure with the radio continuum} \label{sec:disc_radio_cont}

Moving forward, radio continuum observations should prove a valuable tool to identify, and constrain the properties of galaxies experiencing RPS.  LoTSS has already observed a substantial portion of the northern sky; however, the survey is not yet complete.  As the rest of the northern sky is observed with LOFAR, more and more jellyfish galaxies will be identified in newly observed clusters.  Furthermore, the number of galaxy groups ($M_\mathrm{halo} < 10^{14}\,\mathrm{M_\odot}$) vastly surpasses the number of massive clusters in the sky.  In a future study we plan to similarly survey galaxy groups in LoTSS to search for jellyfish galaxies in these lower mass systems.  Because of the very large areas covered by LoTSS, this will be the most extensive search for RPS in galaxy groups to date.
\par
Previous works \citep[e.g.][]{murphy2009,vollmer2009,chen2020,muller2021} have used radio continuum observations to probe the effects of RPS in clusters.  In particular, the VLA observations of the Coma Cluster from \citet{chen2020} can be directly compared to the LoTSS jellyfish galaxies that we identify in Coma in this work.  \citeauthor{chen2020} identify 10 galaxies with 1.4 GHz radio continuum tails out of a total sample of 20 Coma galaxies.  We also detect radio continuum tails at 144 MHz for the 8 tailed galaxies from \citeauthor{chen2020} that pass our stellar mass completeness cut (GMP 4570 and GMP 4629 fall below the SDSS stellar mass completeness limit at the redshift of Coma).  Furthermore, we also detect radio tails at 144 MHz for IC 3949 and NGC 4853 that are not detected at 1.4 GHz in \citet{chen2020}.  \citeauthor{chen2020} present marginal evidence for steeper spectral indices in the tail regions relative to the galaxy disks.  This could explain why we are able to detect tails at 144 MHz that are not seen at 1.4 GHz.  For the 8 galaxies with tails detected at both frequencies, the VLA and LOFAR images could be used in conjunction to more accurately constrain the spectral indices in these ram pressure stripped tails.  We leave this for a future work since this only comprises a narrow subset of the total jellyfish sample.
\par
Beyond Coma, as more of the LoTSS footprint is also observed with the LOFAR low-band antenna (LBA) via the LOFAR LBA Sky Survey (LoLSS, \citealt{deGasperin2021}), as well as APERTIF and RACS \citep{mcconnell2020}, multi-frequency observations will become available for a number of the jellyfish galaxies identified in this work.    The results of this work are consistent with the cosmic rays observed in these tails mostly originating from star formation within the disk, that are subsequently stripped out of the galaxy.  If this is indeed the case, steeper spectral indices should be observed in the tail regions due to spectral ageing of the stripped electrons.  As the multi-frequency observations become available, spectral index maps can be produced for the jellyfish galaxies presented here and this prediction can be directly tested.  We do note that compared to the HBA, the LBA has a higher noise level (in terms of flux density) and lower resolution ($\sim\!15''$); therefore, such spectral index maps may only be possible for the brightest, most prominent jellyfish galaxies in our sample.  \citet{muller2021} show that the stripped tail in the jellyfish galaxy J0206 has a clear spectral index gradient, with the spectral index, $\alpha$, steepening to $2$ in the tail compared to $0.7$ in the galaxy disk (for $S_\nu \propto \nu^{-\alpha}$).  If steep spectra are a generic feature of RPS tails, then LOFAR is truly an ideal instrument to identify these objects due to its low-frequency coverage.
\par
A related question is how much time needs to elapse after stripping before the stripped plasma has aged sufficiently for a spectral index gradient to be observable (assuming no re-acceleration after stripping)?  From \citet{bruno2019}, the radiative age of such a plasma, with spectral index at $t=0$ of $\Gamma$ and an observed spectral index $\alpha$, measured between frequencies $\nu_1$ and $\nu_2$ (in GHz) is given by
\begin{equation}
    t_\mathrm{rad} = 1590\,\frac{\sqrt{B}}{(B^2+B_\mathrm{CMB}^2)\sqrt{1+z}}\,\sqrt{\frac{(\alpha-\Gamma)\ln (\nu_1/\nu_2)}{\nu_1 - \nu_2}}\;\mathrm{[Myr]},
\end{equation}
\noindent
where $B$ is the magnetic field strength in $\mathrm{\mu G}$, $z$ is the redshift, and $B_\mathrm{CMB}=3.25\,(1+z)^2\;\mathrm{[\mu G]}$ is the equivalent CMB magnetic field strength.  Taking a $\sim\!\mathrm{\mu G}$ magnetic field, $\nu_1=144\,\mathrm{MHz}$, $\nu_2=54\,\mathrm{MHz}$ for the HBA and LBA, and $z=0.035$ (roughly the median for our sample), we can ask what is the radiative age that corresponds to the minimum spectral index difference measurable with HBA and LBA?  If we are sensitive to spectral index differences of $\alpha - \Gamma \gtrsim 0.5$ (for example, $\Gamma \sim 0.5$, $\alpha \gtrsim 1$), this corresponds to a radiative age of $\sim\!200\,\mathrm{Myr}$.  Therefore, if the plasma that we observe in these tails was stripped more than $200\,\mathrm{Myr}$ ago, we should be able to measure a spectral index gradient along the tail with the HBA and LBA, if a simple picture of synchrotron ageing is correct.
\par
One way to roughly estimate the time since stripping for the observed tails, is to assume a stripping speed in conjunction with observed tails lengths.  For an average stripping speed, the travel time between the centre of the galaxy and the far edge of the stripped tail is given by
\begin{equation} \label{eq:ttail}
    t_\mathrm{tail} \simeq 100\,\left(\frac{\ell_\mathrm{tail}}{10\,\mathrm{kpc}}\right)\,\left(\frac{100\,\mathrm{km\,s^{-1}}}{\bar{v}}\right)\;\mathrm{[Myr]},
\end{equation}
\noindent
where $\ell_\mathrm{tail}$ is the length of the observed tail and $\bar{v}$ the average speed at which the plasma is stripped from the galaxy.  We note that the tail lengths that we observe are likely shorter than the intrinsic tails lengths, both due to sensitivity limitations of the observations and projection effects.  Therefore, Equation~\ref{eq:ttail} is really a lower limit on the travel time.  If we assume $\bar{v} \sim 100\,\mathrm{km\,s^{-1}}$ \citep[e.g.][]{tonnesen2014}, for the shortest tails in our sample, $\ell_\mathrm{tail} \sim 10\,\mathrm{kpc}$, the travel time is $\sim\!100\,\mathrm{Myr}$ and therefore spectral index steepening along the tail may be difficult to observe in those objects.  For the longest tails in our sample, $\ell_\mathrm{tail} \sim 50\,\mathrm{kpc}$, the travel time becomes $\sim\!500\,\mathrm{Myr}$.  Thus, the more extreme examples are most promising for measuring any spectral index gradients.  We stress that this argument is highly dependent on the stripping speed, $\bar{v}$, which is not well constrained.  If $\bar{v}$ is closer to $1000\,\mathrm{km\,s^{-1}}$, then the travel times become very short.  Recently, \citet{bellhouse2019} estimate a stripping timescale of $\sim\!0.6-1.2\,\mathrm{Gyr}$ for the GASP jellyfish galaxy JO201.  If such a timescale also holds for the jellyfish galaxies in our sample, then this would imply that measuring spectral index steepening along the tails should be feasible with the HBA and LBA.

\subsection{Simulations of magnetized ram pressure stripping}

Recent work has explored the impact of $\sim\! \mathrm{\mu G}$ magnetic fields on the rate of gas removal via RPS and the structure of RPS tails by comparing purely hydrodynamic (HD) and magneto-hydrodynamic (MHD) simulations.  While the amount of gas removed by RPS is similar in both the HD and MHD case \citep{tonnesen2014}, the structure of the tails differ.  In the MHD case, tails are smoother and narrower, and in some cases filametary, whereas for pure HD simulations stripped tails appear far clumpier \citep{ruszkowski2014,vijayaraghavan2017}.  In the plane of the sky these filamentary tails can appear bifurcated, which is consistent with some observed RPS tails in nearby cluster galaxies \citep[e.g.][]{sun2006,sun2010}.
\par
The simple fact that we observe jellyfish tails through synchrotron emission is clear evidence that many (perhaps all) RPS tails are magnetized.  With a $6''$ beam we are not able to resolve structure within the observed tails; therefore, we cannot comment on whether the predicted filamentary, bifurcated structure is present, or even whether the tails are smooth or clumpy.  That said, by incorporating the international LOFAR baselines it is possible for LOFAR to reach subarcsecond resolution at 144 MHz \citep{varenius2015,ramirez-olivencia2018,sweijen2021}.  If the sensitivity at subarcsecond resolutions remains high enough to detect these stripped tails, such features will be highly resolved and it will be possible to comment on the tail morphologies.  We do stress that this will only be possible for tails that are very bright at 144 MHz, due to the large amount of extended flux that is resolved out by the long baselines.  Therefore this would likely be limited to some of the most extreme examples.
\par
\citet{ramos-martinez2018} also perform wind tunnel simulations of RPS, both for a pure HD case and an MHD case.  \citeauthor{ramos-martinez2018} reach similar conclusions regarding the smoother tail morphology when including MHD compared to a clumpier tail in the pure HD case.  They also show that the inclusion of magnetic fields leads to a strongly flared disk, which results in oblique shocks at the interaction interface driving increased gas densities and gas inflows towards the centre of the galaxy. These gas inflows are clear in the MHD case, but far weaker in the purely HD run.  \citeauthor{ramos-martinez2018} speculate that this should lead to episodes of enhanced star formation for galaxies undergoing RPS, which is fully consistent with the enhanced SFRs that we report in this work (Fig.~\ref{fig:SFMS}).

\subsection{Prospects for automated selection of LoTSS jellyfish galaxies} \label{sec:disc_auto_select}

\begin{figure}
    \centering
    \includegraphics[width=0.9\columnwidth]{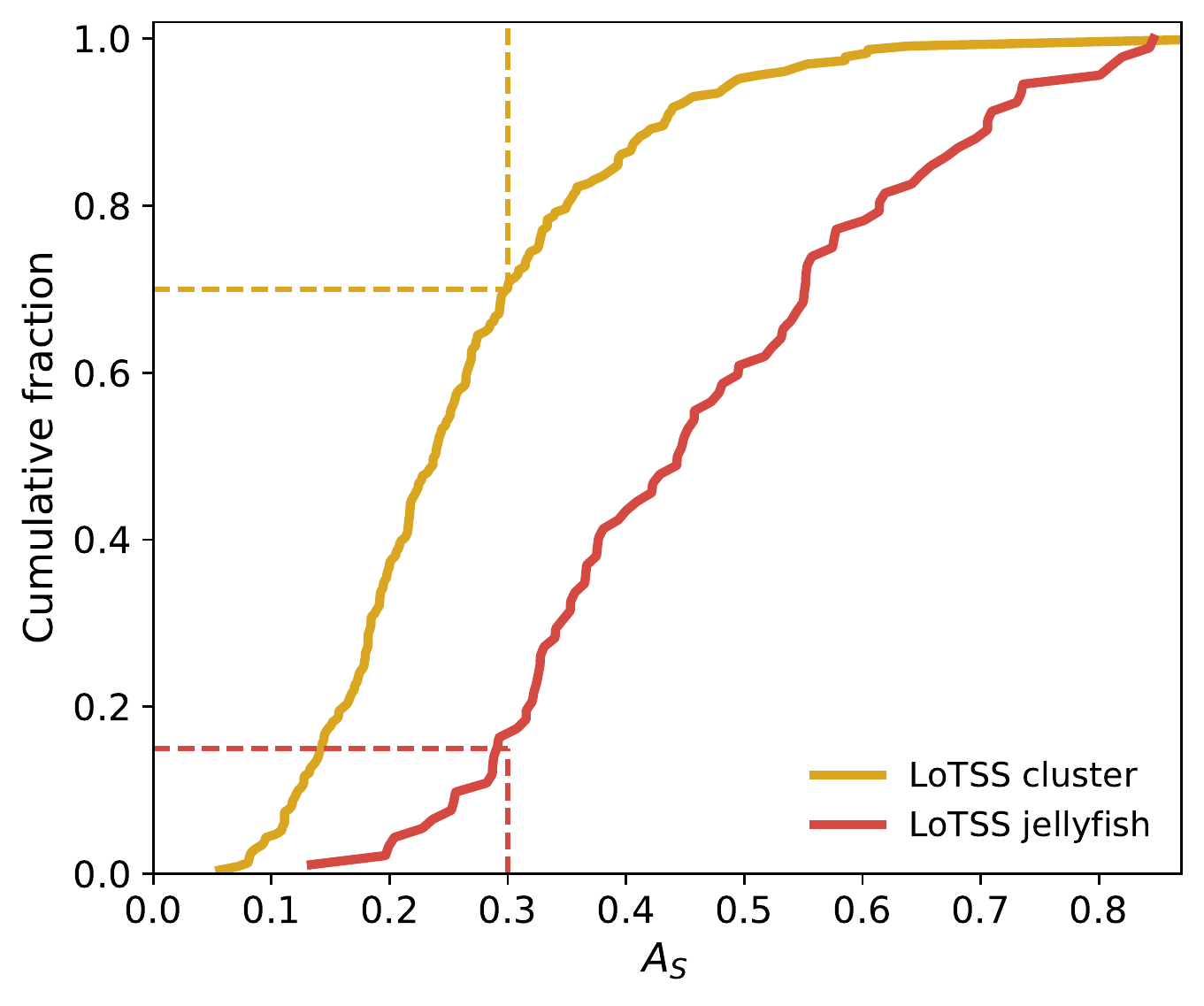}
    \caption{Empirical distribution functions of 144 MHz shape asymmetry for LoTSS cluster galaxies (gold) and LoTSS jellyfish galaxies (red).  High asymmetries correspond to larger $A_S$.}
    \label{fig:auto_select}
\end{figure}

As mentioned in Sect.~\ref{sec:disc_radio_cont}, the large fraction of the SDSS spectroscopic footprint that has been imaged by LoTSS makes this an ideal survey to search for jellyfish galaxies in lower mass SDSS groups.  Groups are an important regime for studies of environmental quenching since the majority of galaxies in the local Universe reside in galaxy groups, and previous works have shown that star formation quenching in groups prior to accretion onto galaxy clusters (known as `pre-processing') may account for a substantial portion of the cluster red sequence \citep[e.g.][]{fujita2004,mcgee2009,vonderlinden2010,haines2015,roberts2017,pallero2020}.  There are $\sim$10 times more galaxies in groups than clusters at low-$z$ \citep[e.g.][]{eke2005}; therefore, for such a search the volume of galaxies requiring visual identification of RPS tails will quickly become overwhelming.  It would be ideal if instead jellyfish galaxies in LoTSS could be selected by an automated algorithm, or at the very least, that an automated process could flag a subset of jellyfish candidates that could then be reasonably checked by-eye for confirmation.  A full exploration of this subject is beyond the scope of this paper; however, below we will discuss one simple scheme that can automatically select LoTSS jellyfish galaxies reasonably well.
\par
The sample of by-eye identified jellyfish galaxies from this work is a valuable set for testing any potential automated classification schemes (ie.\ how well do automated selections reproduce the by-eye identifications from this work?).  We suggest the `shape asymmetry' ($A_S$, \citealt{pawlik2016}) of the 144 MHz segmentation maps, as described in Sect.~\ref{sec:144asym}, as a potentially useful selection tool for LoTSS jellyfish galaxies.  In Fig.~\ref{fig:auto_select} we plot EDFs of the 144 MHz shape asymmetry for both LoTSS cluster galaxies (gold) and LoTSS jellyfish galaxies (red).  There is clear separation between the LoTSS cluster galaxies and LoTSS jellyfish, with the jellyfish galaxies shifted to large shape asymmetries.  For example, selecting sources with $A_S>0.3$ only excludes $\sim$15\% of LoTSS jellyfish galaxies but excludes 70\% of LoTSS cluster galaxies.  We stress that we only include star-forming galaxies, which removes much of the potential contamination from bent AGN jets (e.g. head-tail radio galaxies).  Such a simple selection does not produce a pure sample of jellyfish galaxies, but it is effective at pruning the large sample to a much smaller subset of `jellyfish candidates' for visual inspection.
\par
We show this simple example for illustrative purposes, but a more sophisticated selection based on multiple parameters could likely do even better.  Furthermore, this problem may lend itself well to machine learning techniques.  The sample of LoTSS jellyfish in this work is still fairly small as a training set; however, as
the search for LoTSS jellyfish galaxies extends to more groups
and clusters and the sample size increases, this may become feasible.

\subsection{Summary} \label{sec:summary}

In this paper we use the high sensitivity, wide-field, and high-resolution observations from the LOFAR Two-metre Sky Survey to identify 95 jellyfish galaxies, across 21 low-$z$ clusters, with stripped radio tails at 144 MHz.  The primary results of this work are the following:
\begin{enumerate}
    \itemsep0.5em
    \item The cosmic rays observed in the tails of LoTSS jellyfish are consistent with primarily originating from star formation in the disk. There is no evidence for widespread AGN activity in this sample of jellyfish galaxies.

    \item After carefully controlling for redshift-dependent selection effects, with find no evidence for strong variations in the frequency of jellyfish galaxies from cluster to cluster.

    \item In projected phase space, jellyfish galaxies are found preferentially at small cluster-centric radius and large velocity offsets.  Relative to normal star-forming cluster galaxies, jellyfish galaxies are 2-3 times more likely to occupy this phase space region.

    \item The asymmetries of 144 MHz tails, relative to the optical centre of the galaxy, are largest for jellyfish galaxies at small cluster-centric radii.

    \item Observed 144 MHz tails are systematically oriented away from the cluster centre, suggesting that most jellyfish galaxies are being stripped on their first infall.

    \item LoTSS jellyfish galaxies show enhanced star formation rates.  This is true relative to the field star-forming main sequence, relative to normal star-forming cluster galaxies, and relative to LoTSS-detected cluster galaxies without radio tails.

    \item LoTSS cluster galaxies fall along the normal $L_\mathrm{144MHz} - \mathrm{SFR}$ relation for low-$z$ galaxies published by \citet{gurkan2018}.  The $L_\mathrm{144MHz} - \mathrm{SFR}$ relation for LoTSS jellyfish galaxies has a shallower slope and larger normalization compared to the \citet{gurkan2018} relation.
\end{enumerate}
\noindent
This study is the first search for jellyfish galaxies in a wide-area radio survey.  In this initial work we have focused on low redshift clusters; however, moving forward we will extend this approach to lower mass groups as well as higher redshifts.  These methods have proven effective at identifying a large number of jellyfish galaxies, and such a sample will be valuable for constraining the properties of these extreme objects moving forward.

\begin{acknowledgements}
The authors thank the anonymous referee for their helpful suggestions and Bianca Poggianti for providing comments on an earlier draft.  IDR and RJvW acknowledge support from the ERC Starting Grant Cluster Web 804208. SLM acknowledges support from STFC through grant number ST/N021702/1.  AB acknowledges support from the VIDI research programme with project number 639.042.729, which is financed by the Netherlands Organisation for Scientific Research (NWO).
AD acknowledges support by the BMBF Verbundforschung under the grant 05A20STA. AI acknowledges the Italian PRIN-Miur 2017 (PI A. Cimatti).
\par
This paper is based on data obtained with the International LOFAR Telescope (ILT). LOFAR \citep{vanhaarlem2013} is the LOw Frequency ARray designed and constructed by ASTRON. It has observing, data processing, and data storage facilities in several countries, which are owned by various parties (each with their own funding sources) and are collectively operated by the ILT foundation under a joint scientific policy. The ILT resources have benefited from the following recent major funding sources: CNRS-INSU, Observatoire de Paris and Universit\'e d'Orl\'eans, France; BMBF, MIWF-NRW, MPG, Germany; Science Foundation Ireland (SFI), Department of Business, Enterprise and Innovation (DBEI), Ireland; NWO, The Netherlands; The Science and Technology Facilities Council, UK; Ministry of Science and Higher Education, Poland; The Istituto Nazionale di Astrofisica (INAF), Italy. This research made use of the Dutch national e-infrastructure with support of the SURF Cooperative (e-infra 180169) and the LOFAR e-infra group. The J\"ulich LOFAR Long Term Archive and the GermanLOFAR network are both coordinated and operated by the J\"ulich Supercomputing Centre (JSC), and computing resources on the supercomputer JUWELS at JSC were provided by the Gauss Centre for Supercomputinge.V. (grant CHTB00) through the John von Neumann Institute for Computing (NIC). This research made use of the University of Hertfordshire high-performance computing facility (\url{http://uhhpc. herts.ac.uk}) and the LOFAR-UK computing facility located at the University of Hertfordshire and supported by STFC [ST/P000096/1], and of the Italian LOFAR IT computing infrastructure supported and operated by INAF, and by the Physics Department of Turin University (under an agreement with Consorzio Interuniversitario per la Fisica Spaziale) at the C3S Supercomputing Centre, Italy.
\par
The Jülich LOFAR Long Term Archive and the German LOFAR network are both coordinated and operated by the Jülich Supercomputing Centre (JSC), and computing resources on the supercomputer JUWELS at JSC were provided by the Gauss Centre for Supercomputing e.V. (grant CHTB00) through the John von Neumann Institute for Computing (NIC).
\end{acknowledgements}

%
%

\bibliographystyle{aa}
\bibliography{main}

\begin{appendix}
\onecolumn
\section{Jellyfish galaxies} \label{sec:img_appendix}

\begin{figure}
    \centering
    \includegraphics[width=\textwidth]{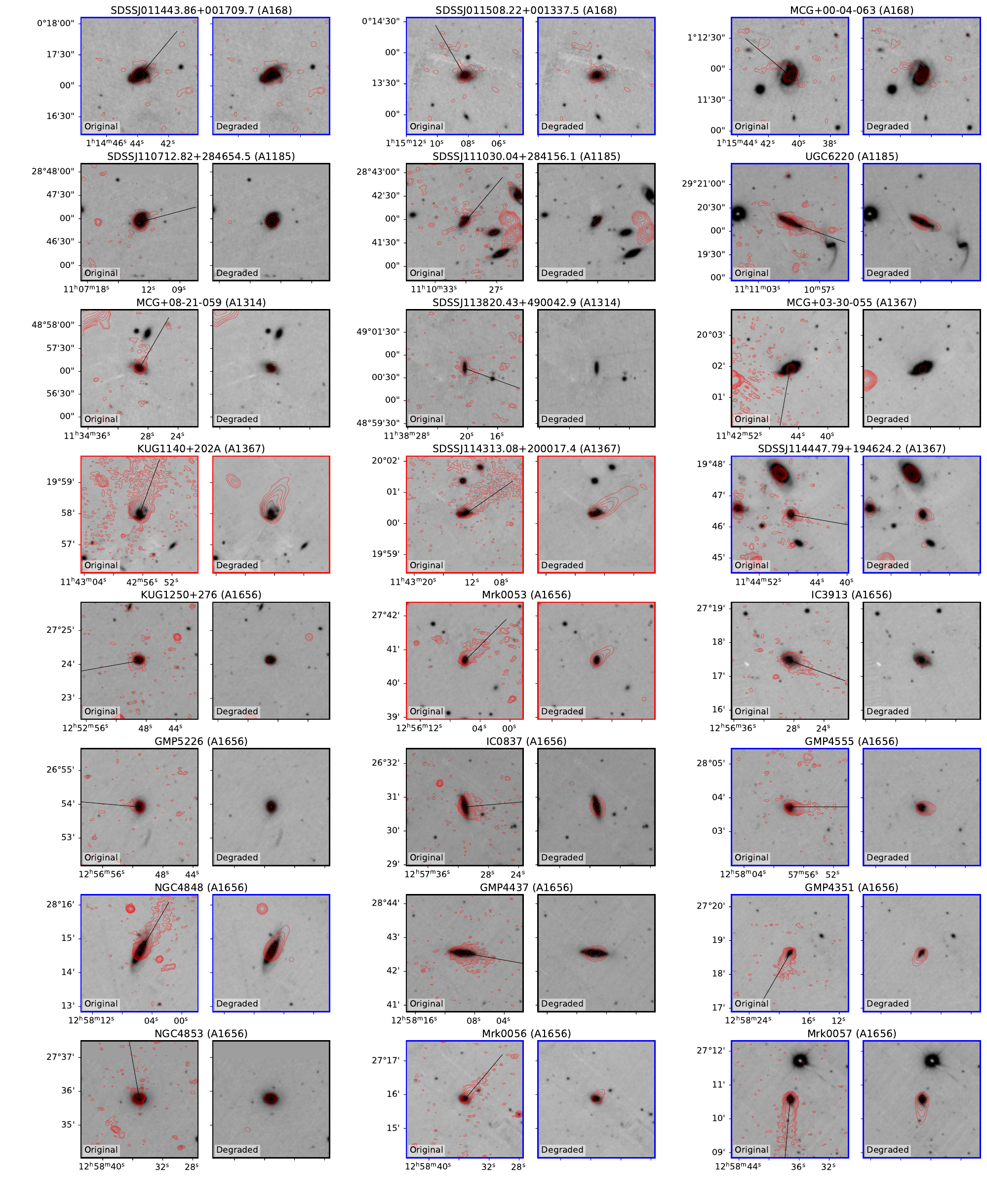}
    \caption{$100 \times 100\,\mathrm{kpc}$ Pan-STARRs $g$-band, LOFAR 144 MHz overlay images for LoTSS jellyfish galaxies.  144 MHz contours are logarithmically spaced starting at $2 \times$ the rms.  The estimated direction of the stripped tail is also marked for each galaxy.  For each galaxy the original LoTSS contours (left) and the degraded LoTSS contours (right, see text) are shown.  Galaxies classified as jellyfish according to both the original and the degraded LoTSS images are highlighted in blue.  Galaxies classified as jellyfish according to both the original and the degraded LoTSS images, but with $M_\mathrm{star} < 10^{9.7}\,\mathrm{M_\odot}$, are highlighted in red.}
    \label{fig:jellyfish_panels1}
\end{figure}

\begin{figure}
    \centering
    \includegraphics[width=\textwidth]{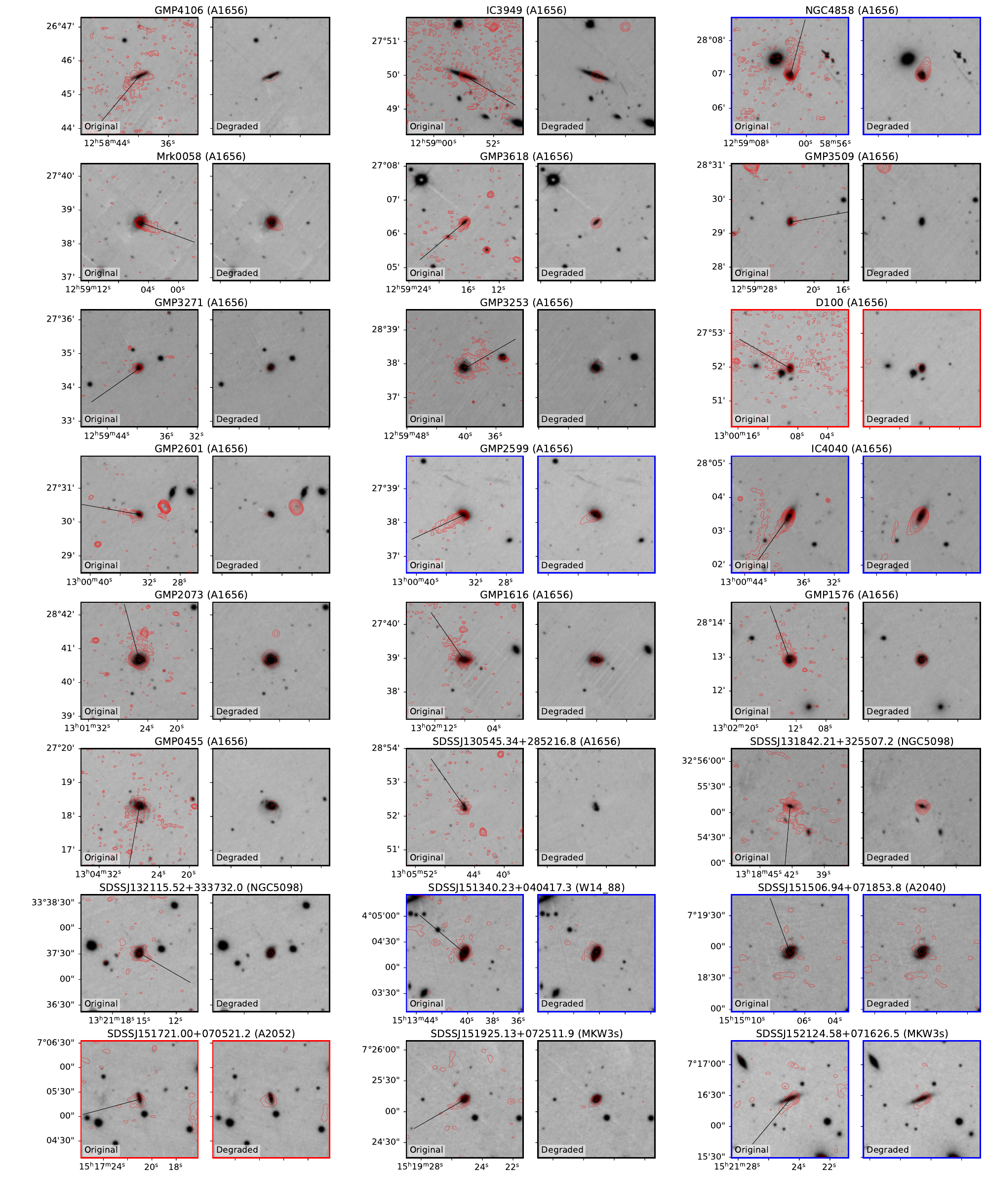}
    \caption{Jellyfish galaxy overlay images, continued.}
    \label{fig:jellyfish_panels2}
\end{figure}

\begin{figure}
    \centering
    \includegraphics[width=\textwidth]{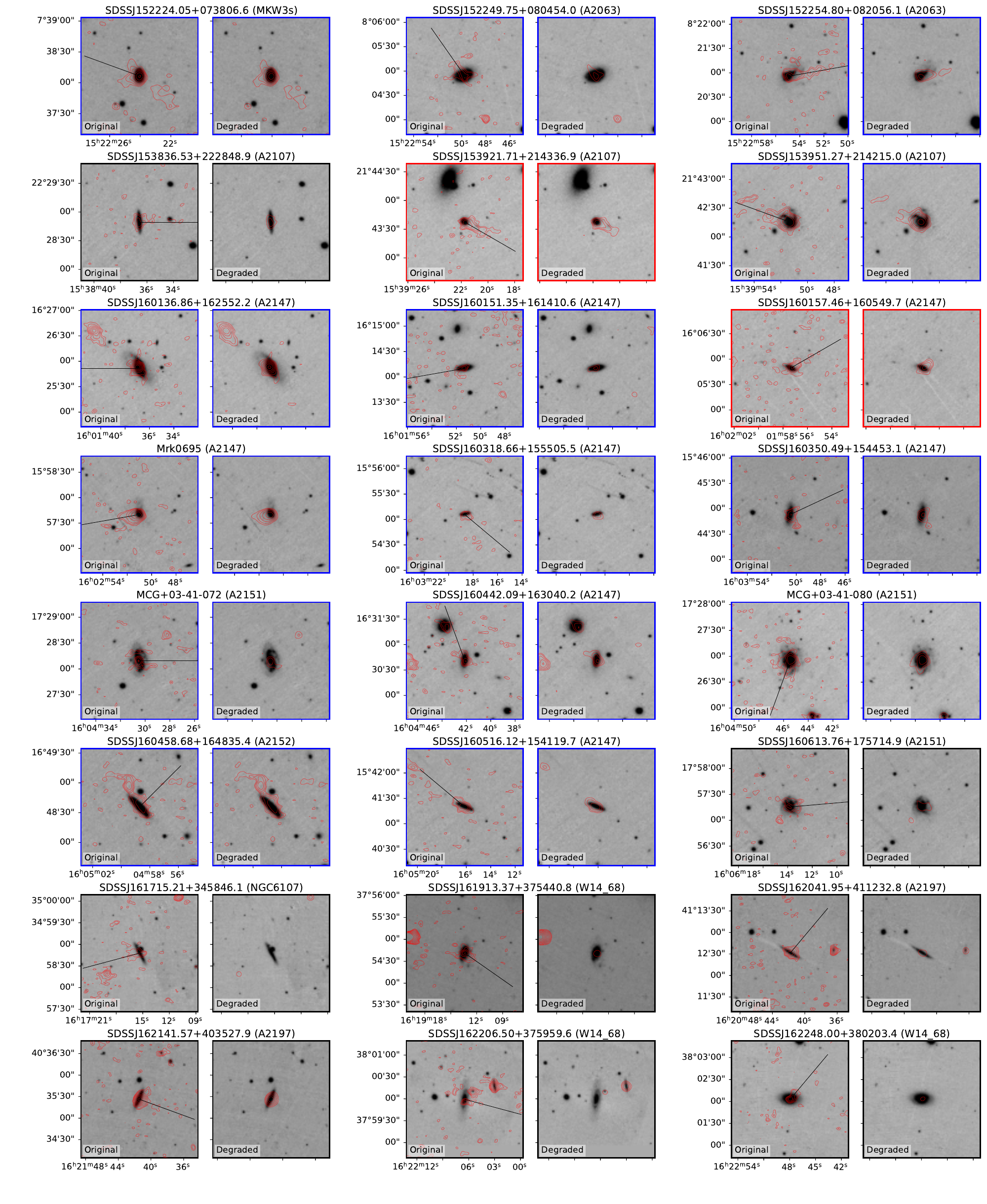}
    \caption{Jellyfish galaxy overlay images, continued.}
    \label{fig:jellyfish_panels3}
\end{figure}

\begin{figure}
    \centering
    \includegraphics[width=\textwidth]{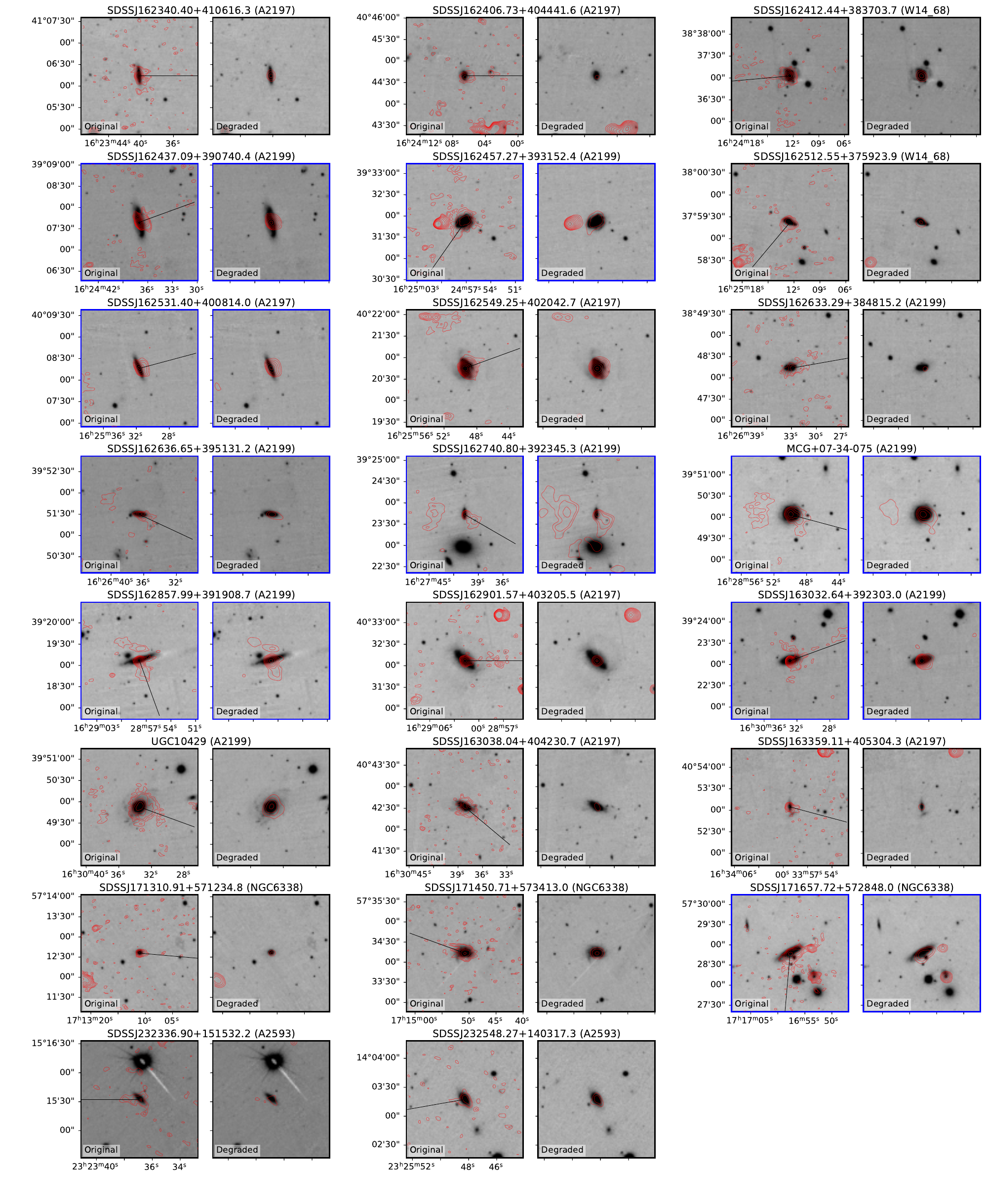}
    \caption{Jellyfish galaxy overlay images, continued.}
    \label{fig:jellyfish_panels4}
\end{figure}

In this appendix we list all LoTSS jellyfish galaxies in Table~\ref{tab:jellyfish_sample}, we also show $100 \times 100\,\mathrm{kpc}$ Pan-STARRs $g$-band images with LOFAR 144 MHz contours overlaid for each jellyfish galaxy in Figs~\ref{fig:jellyfish_panels1}-\ref{fig:jellyfish_panels4}.  We show both the original LoTSS contours as well as the `degraded' LoTSS contours described in Sect.~\ref{sec:cluster_fractions}.  Panels that are highlighted in blue correspond to galaxies that are identified as jellyfish galaxies with both the original LoTSS imaging and the degraded LoTSS imaging. Panels highlighted in red correspond to galaxies identified as jellyfish galaxies with both the original LoTSS imaging and the degraded LoTSS imaging but with stellar masses $<10^{9.7}\,\mathrm{M_\odot}$.  For three galaxies, IC4040, D100, and SDSSJ114313.08+200017.4, the lowest contour corresponds to $3\times$ the rms instead of $2\times$ the rms.  This is for visibility reasons as those galaxies are spatially coincident with diffuse radio sources in their host clusters that can obscure the tail.

\input{RPtable}

\end{appendix}

\end{document}

%% file: RPtable.tex
\begin{ThreePartTable}
\begin{TableNotes}
\footnotesize
\item [a] \citet{salim2016,salim2018}
\item [b] 144 MHz flux density, see text for details
\end{TableNotes}
\begin{longtable}{l c c c c c c c c c}
\caption{LoTSS jellyfish galaxies} \\
\toprule
Name & Cluster & RA & Dec & $z$ & $\log M_\mathrm{star}$\tnote{a} & $\log \mathrm{SFR}$\tnote{a} & $R/R_{180}$ & $\Delta v / \sigma$ & $f_\mathrm{144}$\tnote{b} \\
& & [deg] & [deg] & & [$\mathrm{M_\odot}$] & [$\mathrm{M_\odot\,yr^{-1}}$] & & & [mJy] \\
\midrule
\endfirsthead
\toprule
Name & Cluster & RA & Dec & $z$ & $\log M_\mathrm{star}$\tnote{a} & $\log \mathrm{SFR}$\tnote{a} & $R/R_{180}$ & $\Delta v / \sigma$ & $f_\mathrm{144}$\tnote{b} \\
& & [deg] & [deg] & & [$\mathrm{M_\odot}$] & [$\mathrm{M_\odot\,yr^{-1}}$] & & & [mJy] \\
\midrule
\endhead
\midrule
\multicolumn{5}{r}{{Continued on next page}} \\
\bottomrule
\endfoot
\bottomrule
\insertTableNotes
\endlastfoot
SDSSJ011443.86+001709.7 & A168 & 18.6828 & 0.2860 & 0.0452 & 10.7 & 0.27 & 0.20 & 0.15 & 11.9 \\
MCG+00-04-063 & A168 & 18.9192 & 1.1982 & 0.0442 & 11.0 & 0.55 & 1.00 & 0.37 & 8.8 \\
SDSSJ011508.22+001337.5 & A168 & 18.7843 & 0.2271 & 0.0437 & 10.5 & 0.26 & 0.26 & 0.64 & 9.1 \\
SDSSJ110712.82+284654.5 & A1185 & 166.8035 & 28.7818 & 0.0316 & 10.0 & 0.26 & 0.70 & 0.94 & 8.8 \\
SDSSJ111030.04+284156.1 & A1185 & 167.6252 & 28.6989 & 0.0348 & 10.3 & -0.59 & 0.04 & 0.68 & 3.4 \\
UGC6220 & A1185 & 167.7493 & 29.3365 & 0.0292 & 11.0 & 0.03 & 0.59 & 2.14 & 28.5 \\
MCG+08-21-059 & A1314 & 173.6213 & 48.9510 & 0.0282 & 9.5 & -0.15 & 0.16 & 2.88 & 5.7 \\
SDSSJ113820.43+490042.9 & A1314 & 174.5851 & 49.0120 & 0.0303 & 9.7 & -0.51 & 0.67 & 1.51 & 1.7 \\
SDSSJ114447.79+194624.2 & A1367 & 176.1992 & 19.7734 & 0.0282 & 9.8 & 0.20 & 0.10 & 2.96 & 29.8 \\
MCG+03-30-055 & A1367 & 175.6882 & 20.0323 & 0.0211 & 10.6 & 0.30 & 0.39 & 0.38 & 5.2 \\
KUG1140+202A & A1367 & 175.7352 & 19.9662 & 0.0243 & 9.7 & 0.01 & 0.34 & 1.12 & 133.6 \\
SDSSJ114313.08+200017.4 & A1367 & 175.8046 & 20.0048 & 0.0234 & 9.4 & 0.15 & 0.33 & 0.70 & 74.2 \\
IC3949 & A1656 & 194.7332 & 27.8334 & 0.0251 & 10.6 & -0.16 & 0.10 & 0.49 & 14.0 \\
NGC4858 & A1656 & 194.7586 & 28.1157 & 0.0314 & 10.2 & 0.56 & 0.10 & 2.96 & 50.6 \\
Mrk0058 & A1656 & 194.7721 & 27.6444 & 0.0181 & 9.8 & -0.04 & 0.17 & 2.26 & 25.0 \\
GMP3618 & A1656 & 194.8195 & 27.1061 & 0.0280 & 10.1 & 0.09 & 0.43 & 1.63 & 12.0 \\
GMP3509 & A1656 & 194.8465 & 28.4886 & 0.0233 & 9.6 & -0.50 & 0.27 & 0.22 & 2.8 \\
GMP3271 & A1656 & 194.9159 & 27.5765 & 0.0167 & 9.1 & -0.66 & 0.19 & 2.81 & 4.2 \\
GMP1576 & A1656 & 195.5533 & 28.2148 & 0.0273 & 10.0 & 0.18 & 0.32 & 1.35 & 19.8 \\
D100 & A1656 & 195.0381 & 27.8665 & 0.0177 & 9.3 & -0.08 & 0.08 & 2.42 & 7.5 \\
GMP2601 & A1656 & 195.1398 & 27.5041 & 0.0186 & 9.0 & -0.54 & 0.25 & 2.06 & 5.2 \\
GMP4106 & A1656 & 194.6665 & 26.7595 & 0.0249 & 9.1 & -0.66 & 0.61 & 0.41 & 11.3 \\
GMP2599 & A1656 & 195.1403 & 27.6378 & 0.0250 & 9.8 & 0.06 & 0.19 & 0.45 & 25.1 \\
IC4040 & A1656 & 195.1578 & 28.0581 & 0.0255 & 10.3 & 0.64 & 0.12 & 0.65 & 101.7 \\
GMP3253 & A1656 & 194.9172 & 28.6308 & 0.0178 & 9.4 & -0.32 & 0.34 & 2.38 & 19.1 \\
Mrk0057 & A1656 & 194.6553 & 27.1766 & 0.0256 & 9.8 & 0.25 & 0.41 & 0.69 & 43.6 \\
NGC4848 & A1656 & 194.5233 & 28.2426 & 0.0240 & 10.8 & 0.99 & 0.22 & 0.06 & 127.8 \\
NGC4853 & A1656 & 194.6466 & 27.5964 & 0.0256 & 10.8 & 0.41 & 0.21 & 0.69 & 15.9 \\
GMP4351 & A1656 & 194.5776 & 27.3108 & 0.0247 & 9.7 & -0.09 & 0.35 & 0.33 & 29.3 \\
GMP4437 & A1656 & 194.5385 & 28.7086 & 0.0254 & 10.4 & 0.15 & 0.41 & 0.61 & 40.2 \\
GMP2073 & A1656 & 195.3545 & 28.6772 & 0.0292 & 10.3 & 0.32 & 0.41 & 2.10 & 25.6 \\
GMP4555 & A1656 & 194.4905 & 28.0618 & 0.0271 & 9.9 & 0.07 & 0.19 & 1.27 & 38.3 \\
IC0837 & A1656 & 194.3800 & 26.5122 & 0.0241 & 10.3 & 0.34 & 0.76 & 0.10 & 39.8 \\
GMP5226 & A1656 & 194.2132 & 26.8989 & 0.0208 & 10.4 & 0.01 & 0.61 & 1.20 & 9.3 \\
IC3913 & A1656 & 194.1191 & 27.2913 & 0.0251 & 10.0 & -0.12 & 0.48 & 0.49 & 17.9 \\
Mrk0053 & A1656 & 194.0254 & 27.6781 & 0.0165 & 9.2 & 0.04 & 0.41 & 2.89 & 39.2 \\
KUG1250+276 & A1656 & 193.2037 & 27.4019 & 0.0258 & 9.8 & 0.14 & 0.80 & 0.76 & 10.0 \\
GMP0455 & A1656 & 196.1106 & 27.3043 & 0.0184 & 9.3 & -0.27 & 0.63 & 2.14 & 11.2 \\
SDSSJ130545.34+285216.8 & A1656 & 196.4390 & 28.8713 & 0.0266 & 9.2 & -0.50 & 0.82 & 1.08 & 4.1 \\
Mrk0056 & A1656 & 194.6472 & 27.2647 & 0.0245 & 9.8 & 0.24 & 0.37 & 0.25 & 19.2 \\
GMP1616 & A1656 & 195.5328 & 27.6483 & 0.0230 & 10.2 & 0.42 & 0.32 & 0.34 & 21.1 \\
SDSSJ151506.94+071853.8 & A2040 & 228.7789 & 7.3150 & 0.0433 & 10.4 & 0.29 & 0.64 & 0.82 & 3.6 \\
SDSSJ151721.00+070521.2 & A2052 & 229.3375 & 7.0891 & 0.0321 & 9.5 & -0.34 & 0.17 & 1.59 & 6.1 \\
SDSSJ152249.75+080454.0 & A2063 & 230.7073 & 8.0817 & 0.0328 & 10.5 & -0.14 & 0.55 & 1.13 & 6.1 \\
SDSSJ152254.80+082056.1 & A2063 & 230.7284 & 8.3489 & 0.0311 & 9.7 & 0.03 & 0.28 & 2.07 & 10.8 \\
SDSSJ153836.53+222848.9 & A2107 & 234.6522 & 22.4803 & 0.0390 & 10.4 & -0.12 & 0.90 & 1.33 & 3.4 \\
SDSSJ153921.71+214336.9 & A2107 & 234.8405 & 21.7269 & 0.0417 & 9.7 & 0.14 & 0.11 & 0.15 & 9.0 \\
SDSSJ153951.27+214215.0 & A2107 & 234.9637 & 21.7042 & 0.0422 & 10.0 & 0.31 & 0.11 & 0.42 & 18.2 \\
SDSSJ160516.12+154119.7 & A2147 & 241.3172 & 15.6888 & 0.0352 & 10.1 & 0.16 & 0.61 & 0.48 & 7.3 \\
SDSSJ160442.09+163040.2 & A2147 & 241.1754 & 16.5112 & 0.0363 & 10.5 & 0.04 & 0.63 & 0.03 & 6.5 \\
SDSSJ160350.49+154453.1 & A2147 & 240.9604 & 15.7481 & 0.0308 & 10.2 & 0.00 & 0.35 & 2.28 & 4.7 \\
Mrk0695 & A2147 & 240.7124 & 15.9610 & 0.0352 & 9.8 & 0.42 & 0.11 & 0.48 & 17.0 \\
SDSSJ160157.46+160549.7 & A2147 & 240.4895 & 16.0971 & 0.0420 & 9.5 & 0.06 & 0.12 & 2.30 & 7.0 \\
SDSSJ160151.35+161410.6 & A2147 & 240.4640 & 16.2363 & 0.0353 & 10.5 & -0.41 & 0.22 & 0.44 & 1.7 \\
SDSSJ160318.66+155505.5 & A2147 & 240.8278 & 15.9182 & 0.0419 & 10.1 & -0.18 & 0.20 & 2.26 & 2.6 \\
SDSSJ160136.86+162552.2 & A2147 & 240.4036 & 16.4312 & 0.0435 & 10.9 & 0.75 & 0.38 & 2.91 & 21.5 \\
MCG+03-41-072 & A2151 & 241.1268 & 17.4692 & 0.0396 & 10.6 & 0.33 & 0.22 & 1.07 & 3.6 \\
MCG+03-41-080 & A2151 & 241.1893 & 17.4484 & 0.0355 & 10.5 & 0.44 & 0.24 & 0.70 & 10.3 \\
SDSSJ160613.76+175714.9 & A2151 & 241.5574 & 17.9542 & 0.0386 & 10.0 & 0.25 & 0.39 & 0.64 & 4.1 \\
SDSSJ160458.68+164835.4 & A2152 & 241.2445 & 16.8098 & 0.0431 & 10.8 & 0.28 & 0.43 & 0.02 & 12.7 \\
SDSSJ162531.40+400814.0 & A2197 & 246.3809 & 40.1372 & 0.0329 & 10.5 & 0.30 & 0.64 & 1.00 & 23.7 \\
SDSSJ162901.57+403205.5 & A2197 & 247.2566 & 40.5349 & 0.0278 & 10.7 & 0.14 & 0.34 & 1.25 & 6.4 \\
SDSSJ162549.25+402042.7 & A2197 & 246.4553 & 40.3452 & 0.0290 & 10.4 & 0.90 & 0.49 & 0.72 & 57.1 \\
SDSSJ162406.73+404441.6 & A2197 & 246.0281 & 40.7449 & 0.0306 & 9.5 & -0.05 & 0.51 & 0.01 & 2.8 \\
SDSSJ163038.04+404230.7 & A2197 & 247.6586 & 40.7085 & 0.0267 & 9.5 & -0.13 & 0.43 & 1.73 & 5.7 \\
SDSSJ163359.11+405304.3 & A2197 & 248.4963 & 40.8846 & 0.0315 & 9.9 & -0.23 & 0.86 & 0.38 & 2.6 \\
SDSSJ162340.40+410616.3 & A2197 & 245.9184 & 41.1045 & 0.0330 & 10.5 & -0.24 & 0.56 & 1.04 & 6.0 \\
SDSSJ162041.95+411232.8 & A2197 & 245.1748 & 41.2091 & 0.0301 & 10.0 & -0.09 & 0.97 & 0.23 & 3.8 \\
SDSSJ162141.57+403527.9 & A2197 & 245.4232 & 40.5911 & 0.0312 & 10.5 & -0.04 & 0.86 & 0.25 & 15.0 \\
SDSSJ162437.09+390740.4 & A2199 & 246.1546 & 39.1279 & 0.0352 & 10.8 & 0.82 & 0.70 & 2.39 & 29.4 \\
SDSSJ162633.29+384815.2 & A2199 & 246.6387 & 38.8042 & 0.0345 & 9.7 & -0.08 & 0.67 & 2.05 & 4.9 \\
SDSSJ162857.99+391908.7 & A2199 & 247.2416 & 39.3191 & 0.0339 & 10.9 & 0.49 & 0.19 & 1.76 & 32.5 \\
SDSSJ162740.80+392345.3 & A2199 & 246.9200 & 39.3959 & 0.0349 & 10.1 & 0.06 & 0.19 & 2.25 & 14.5 \\
UGC10429 & A2199 & 247.6387 & 39.8307 & 0.0246 & 10.5 & -0.09 & 0.36 & 2.71 & 26.4 \\
SDSSJ162457.27+393152.4 & A2199 & 246.2387 & 39.5312 & 0.0360 & 10.5 & 0.30 & 0.56 & 2.78 & 16.2 \\
SDSSJ163032.64+392303.0 & A2199 & 247.6361 & 39.3842 & 0.0305 & 10.5 & 0.58 & 0.32 & 0.13 & 45.8 \\
SDSSJ162636.65+395131.2 & A2199 & 246.6527 & 39.8587 & 0.0314 & 10.3 & 0.06 & 0.39 & 0.56 & 6.3 \\
MCG+07-34-075 & A2199 & 247.2091 & 39.8351 & 0.0359 & 10.7 & 0.64 & 0.22 & 2.73 & 20.4 \\
SDSSJ232548.27+140317.3 & A2593 & 351.4511 & 14.0548 & 0.0377 & 10.3 & 0.39 & 0.71 & 1.90 & 3.3 \\
SDSSJ232336.90+151532.2 & A2593 & 350.9038 & 15.2590 & 0.0434 & 10.1 & 0.17 & 0.66 & 0.73 & 3.2 \\
SDSSJ152224.05+073806.6 & MKW3s & 230.6002 & 7.6352 & 0.0402 & 10.2 & 0.34 & 0.30 & 2.66 & 14.1 \\
SDSSJ151925.13+072511.9 & MKW3s & 229.8548 & 7.4200 & 0.0427 & 9.9 & 0.20 & 0.88 & 1.19 & 3.5 \\
SDSSJ152124.58+071626.5 & MKW3s & 230.3525 & 7.2740 & 0.0471 & 10.4 & 0.08 & 0.64 & 1.39 & 4.8 \\
SDSSJ132115.52+333732.0 & NGC5098 & 200.3147 & 33.6256 & 0.0389 & 10.0 & 0.23 & 0.68 & 1.49 & 4.0 \\
SDSSJ131842.21+325507.2 & NGC5098 & 199.6759 & 32.9187 & 0.0359 & 10.1 & -0.10 & 0.50 & 0.45 & 15.9 \\
SDSSJ161715.21+345846.1 & NGC6107 & 244.3135 & 34.9796 & 0.0281 & 9.7 & -0.52 & 0.06 & 1.87 & 2.0 \\
SDSSJ171310.91+571234.8 & NGC6338 & 258.2955 & 57.2097 & 0.0281 & 9.7 & 0.15 & 0.41 & 0.36 & 2.8 \\
SDSSJ171450.71+573413.0 & NGC6338 & 258.7112 & 57.5703 & 0.0268 & 10.2 & -0.11 & 0.20 & 1.31 & 10.2 \\
SDSSJ171657.72+572848.0 & NGC6338 & 259.2405 & 57.4800 & 0.0284 & 10.5 & -0.00 & 0.25 & 0.14 & 17.4 \\
SDSSJ162412.44+383703.7 & W14\_68 & 246.0519 & 38.6177 & 0.0344 & 10.6 & 0.62 & 0.87 & 2.22 & 7.3 \\
SDSSJ162248.00+380203.4 & W14\_68 & 245.7000 & 38.0343 & 0.0305 & 10.5 & 0.14 & 0.15 & 0.50 & 2.4 \\
SDSSJ162206.50+375959.6 & W14\_68 & 245.5271 & 37.9999 & 0.0299 & 10.2 & 0.09 & 0.24 & 0.92 & 1.3 \\
SDSSJ161913.37+375440.8 & W14\_68 & 244.8057 & 37.9113 & 0.0304 & 10.3 & 0.28 & 0.90 & 0.57 & 4.0 \\
SDSSJ162512.55+375923.9 & W14\_68 & 246.3022 & 37.9900 & 0.0282 & 9.7 & -0.20 & 0.51 & 2.11 & 4.8 \\
SDSSJ151340.23+040417.3 & W14\_88 & 228.4176 & 4.0715 & 0.0393 & 10.3 & 0.38 & 0.96 & 1.66 & 12.3 \\
\label{tab:jellyfish_sample}
\end{longtable}
\end{ThreePartTable}